\documentclass[a4paper]{article}
\usepackage[utf8]{inputenc}
\usepackage[utf8]{inputenc}
\usepackage{amsmath,pgfmath,pgffor}
\usepackage{graphicx}
\usepackage{subcaption}
\usepackage{float}
\usepackage{fancyhdr}
\usepackage{cite}
\usepackage{xcolor}
\usepackage{amsfonts}
\usepackage{textcomp}
\usepackage{multicol}
\usepackage{makecell}
\usepackage{booktabs}  

\usepackage[a4paper, total={6in, 8.5in}]{geometry} 

\title{Analysis of Super-resolution Single Molecule Localization Microscopy Data: a tutorial}
\author{Mohamadreza Fazel$^{1,2}$ and Michael J. Wester$^{1,3}$\\
$^1$Department of Physics and Astronomy, University of New Mexico,\\ Albuquerque, New Mexico 87106, USA\\
$^2$Center for Biological Physics, Department of Physics, Arizona State University,\\ Tempe, Arizona 85287, USA\\
$^3$Department of Mathematics and Statistics, University of New Mexico,\\ Albuquerque, New Mexico 87106, USA\\
}
\date{}

\begin{document}
\maketitle
\begin{abstract}
The diffraction of light imposes a fundamental limit on the resolution of light microscopes. This limit can be circumvented by creating and exploiting independent behaviors of the sample at length scales below the diffraction limit. In super-resolution single molecule localization microscopy (SMLM), the independence arises from individual fluorescent labels stochastically switching between dark and fluorescent states, which in turn allows
the pinpointing of fluorophores post experimentally using a sequence of acquired sparse image frames. Finally, the resulting list of fluorophore coordinates is utilized to produce high resolution images or to gain quantitative insight into the underlying biological structures. Therefore, image processing and post-processing are essential stages of SMLM.
Here, we review the latest progress on SMLM data processing and post-processing.
\end{abstract}
Author to whom correspondence should be addressed:  mfazel@asu.edu
\section{Introduction}
\label{Sec1_1}
In the beginning of the 20th century, a major difficulty in the examination of living organisms was their transparency, which allows transmission of a large portion of light without effective scattering or attenuation \cite{renz2013Cytometry}. Fluorescence microscopy techniques were therefore developed which overcome this problem. These techniques improve microscopy images by relying on light from fluorescent probes that stain target structures within a sample instead of weak scattered light from the entire sample.
In spite of this accomplishment, fluorescence microscopy still suffers from major issues: 1) light diffraction; 2) inefficient collection of fluorescent light by microscope lenses; 3) the contribution of light from parts of the specimen that are out of the focal point/plane; 4) photodamage to the sample due to high excitation laser intensities or long data acquisition times. 

Here, we discuss the fundamental resolution limit in microscopy images which arises due to issues 1-2. In fluorescence microscopy under ideal conditions, optimal resolution could be obtained if every fluorophore (point source of light) was represented by a point in microscopy images. However, the diffraction of light emitted by a fluorophore and inefficiently collected by a microscope lens (objective) results in an extended intensity pattern called the Airy disk. Furthermore, the actual pattern often differs from the Airy disk due to optical aberrations, and is referred to as a point spread function (PSF) or an impulse response.

\begin{figure}[H]
\centering
\includegraphics[scale=0.93]{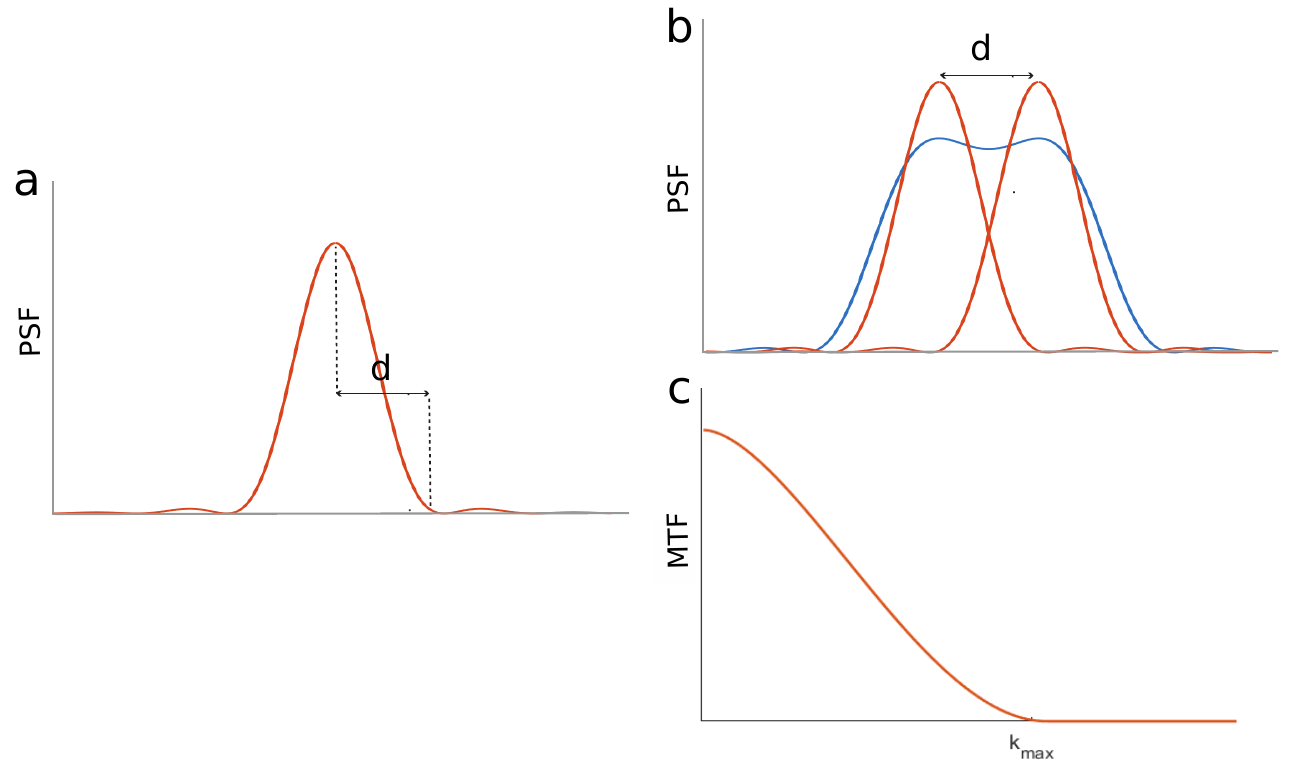}
\caption{Fundamental limit of resolution. (a) 1D diffraction pattern (Airy pattern) from a circular aperture (lens). (b) The red curves are two Airy patterns where their maxima are separated by $d$. The blue curve is the intensity pattern from two noncoherent point sources of light with separation $d$. (c) The absolute value of the Fourier transform of the Airy pattern in panel a. $k_{\mathrm{max}}$ denotes the spatial frequency cutoff, which is the largest spatial frequency passed by the lens.}
\label{Fig_Res}
\end{figure}

\begin{figure}[H]
\centering
\includegraphics[scale=0.9]{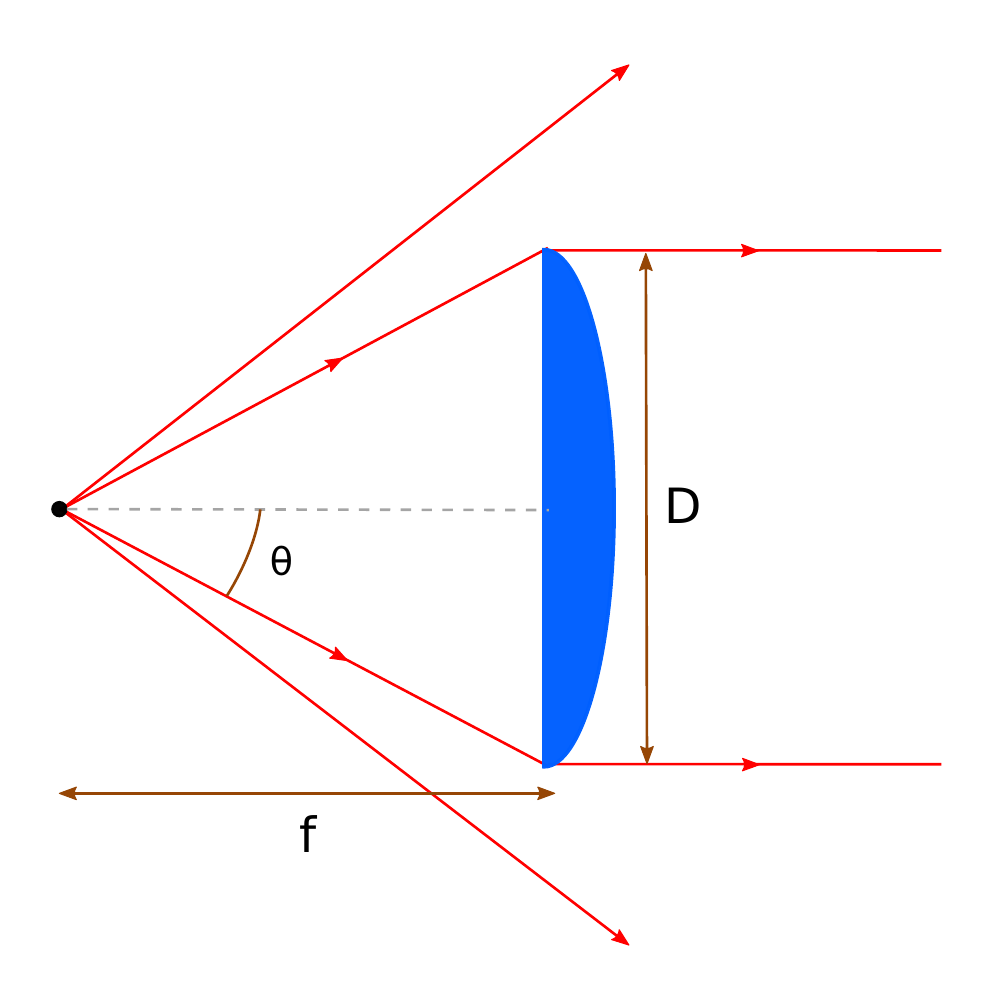}
\caption{Numerical aperture. The numerical aperture ($N_a$) is related to the fraction of light that can be collected by a lens. Here, the lens captures light rays traveling within a cone with half angle $\theta$ given by $\sin\theta \sim D/2f$ for an emitter sitting at the focal point. The numerical aperture is defined by $N_a = n\sin\theta$, where the refractive index $n$ guarantees independence of $N_a$ from the medium by Snell's law.}
\label{NA}
\end{figure}

The PSF plays a central role in determining the resolving power of microscopes. For instance, two-point resolution is defined as the minimum distance, $d$, between two point-like emitters in which their overlapping PSFs can still be distinguished as two individual PSFs by the microscope user \cite{born2013principles}. Rayleigh's criterion for the two-point resolution expresses that two noncoherent sources of light can be resolved when they are at least separated by a distance equal to the separation between the maximum and the first minimum of the diffraction pattern of one of the sources, Fig. \ref{Fig_Res}a-b. Assuming an Airy pattern, the Rayleigh's resolution distance for point sources in a lateral direction is given by \cite{born2013principles,goodman2005introduction}
\begin{equation}
d=0.61\frac{\lambda}{N_a}
\label{Eq1_32}
\end{equation}
where $\lambda$ is the wavelength of the light. $N_a$ is the numerical aperture, which quantifies the efficiency of an objective lens in collecting light, Fig. \ref{NA}. 

On the other hand, the Abbe criterion provides a definition for resolution based on the notion of the optical transfer function (OTF), which is simply the Fourier transform of the PSF. The absolute value of the OTF is called the modulation transfer function (MTF), which gradually decays with increasing spatial frequency, going to zero at a maximum frequency, $k_{\mathrm{max}}$, Fig. \ref{Fig_Res}c. This frequency cutoff puts a fundamental limit on the ability of microscopes (or any imaging system) to collect information, \textit{i.e.}, spatial frequencies, from the sample \cite{stallinga2021optimal}. This effect is called the Abbe diffraction limit and results in the following lateral resolution \cite{born2013principles} 
\begin{equation}
d = \frac{\lambda}{2N_a}
\label{Eq1_1}
\end{equation}
which is related to the inverse of $k_{\mathrm{max}}$. For visible light, $\lambda \sim 550$ nm, and for a typical objective, $N_a \sim 1.3$, so the resolution limit is approximately $200$ nm. 

We now proceed with a short description of resolution along the axial ($z$) direction. Similar to the lateral resolution, the Abbe criterion can be used to find the axial resolution. However, the larger size of the PSF in the axial direction yields a smaller frequency cutoff compared to the lateral case. As a result, the inverse of this cutoff leads to a larger (worse) resolution distance, which can be approximated by \cite{born2013principles,pawley2010handbook}
\begin{equation}
d\sim\frac{2n\lambda}{(N_a)^2},
\end{equation}
where $n$ is the index of refraction.

Now that we have introduced the fundamental limit of resolution as a consequence of issues 1-2, we turn to issues 3-4. Although issues 3-4 do not yield any fundamental limits, they often result in degradation in the quality of microscope images. For example, undesired light from out-of-focus structures stained by fluorophores, or autofluorescence from these structures, contribute to the fluorescence light captured by the objective lens. Moreover, some fluorescence microscopy techniques require high laser intensities or long exposure times that might damage the sample.

To alleviate the issues associated with fluorescence microscopy, different illumination and light collection techniques \cite{marvin1961USPatent,axelrod1981JCB,denk1990Science,hell1992OpticsLetters,bailey1993Nature,voie1993Microscopy} were developed to reduce out-of-focus light and collect fluorescence emission more efficiently. These techniques can be divided into two main categories: point scanning and wide-field microscopes. The point scanning microscopes, such as confocal \cite{marvin1961USPatent} and 4-pi \cite{hell1992OpticsLetters} microscopes, scan the specimen by illuminating a single spot at a time. The main benefit of these techniques is the reduction of out-of-focus light by embedding pinholes in the optical setup that only allow transmission of the in-focus light. In addition to pinholes, the 4-pi technique also uses two objective lenses to collect fluorescent light, yielding a larger numerical aperture which, in turn, results in a higher resolution. In wide-field microscopes, such as the total internal reflection fluorescence (TIRF) microscope \cite{axelrod1981JCB}, the entire specimen is exposed to an exciting laser, as opposed to scanning the sample point by point, allowing for faster data acquisition.

Different fluorescence microscopy techniques pushed the Abbe diffraction barrier to its very limits, however, it was not until the end of 20th century that scientists were able to overcome this barrier and achieve resolutions better than the diffraction limit \cite{hell1994OpticsLetter}.
Such techniques that surpass the diffraction limit and achieve sub-diffraction resolutions (resolutions not restricted by the cutoff in the frequency transmission, $k_{\mathrm{max}}$, by an objective lens) are called super-resolution microscopy or nanoscopy \cite{huang2009super,fornasiero2015super,eggeling2015lens,turkowyd2016single,sahl2017fluorescence,schermelleh2019super,vangindertael2018introduction}. These techniques accomplish sub-diffraction resolution by creating independent behavior for individual fluorophores at scales below the diffraction limit.  
Many super-resolution procedures use the reversible switching property of some fluorescent probes transitioning between a fluorescent state (emitting state) and a dark state (non-emitting state) to obtain sub-diffraction resolution.
These approaches can be classified into two different groups based on how the probes are switched between the dark and fluorescent states: targeted and stochastic switching procedures. The microscopy techniques under the first category are STimulated Emission Depletion (STED) microscopy \cite{hell1994OpticsLetter,klar1999subdiffraction}, Ground State Depletion (GSD) microscopy \cite{hell1995ground}, REversible Saturable Optically Linear Fluorescence Transition (RESOLFT) microscopy \cite{hell2007far} and Saturated Structured Illumination Microscopy (SSIM) \cite{gustafsson2005nonlinear}. These techniques utilize a laser to excite fluorophores within a region of interest (ROI). Moreover, a second laser is used to deterministically switch off the fluorophores in the diffraction limited vicinity of the target fluorophore to accomplish sub-diffraction resolution. On the other hand, Stochastic Optical Reconstruction Microscopy (STORM) \cite{rust2006sub}, direct Stochastic Optical Reconstruction Microscopy (dSTORM) \cite{van2011direct}, Photoactivated Localization Microscopy (PALM) \cite{betzig2006imaging}, Fluorescence Photoactivation Localization Microscopy (FPALM) \cite{hess2006ultra}, Super-resolution Optical Fluctuation Imaging (SOFI) \cite{dertinger2009fast}, DNA-Point Accumulation for Imaging in Nanoscale Topography (DNA-PAINT) \cite{schnitzbauer2017super,pallikkuth2018sequential}, and other variants introduced later \cite{van2011direct,schnitzbauer2017super,pallikkuth2018sequential,lidke2005superresolution,lagerholm2006analysis,egner2007fluorescence,churchman2005single,bates2007multicolor,low2011erbb1,gordon2004single,qu2004nanometer,sharonov2006wide} can be categorized as stochastic switching techniques. These techniques often make use of a wide-field microscopy method, such as TIRF, to activate a small random subset of fluorophores to avoid simultaneous fluorescent light from more than one fluorophore in a diffraction limited region, which in turn allows localization of these fluorophores with a precision better than the diffraction limit.

Most of the stochastic switching super-resolution techniques include an image processing stage to localize fluorophores and are called single molecule localization microscopy (SMLM), or just simply localization microscopy. In what follows, we first discuss the SMLM techniques in more depth and next focus on the processing and post-processing of the data produced from these techniques.
\section{Single Molecule Localization Microscopy}

\begin{figure}[H]
\centering
\includegraphics[scale=1.125]{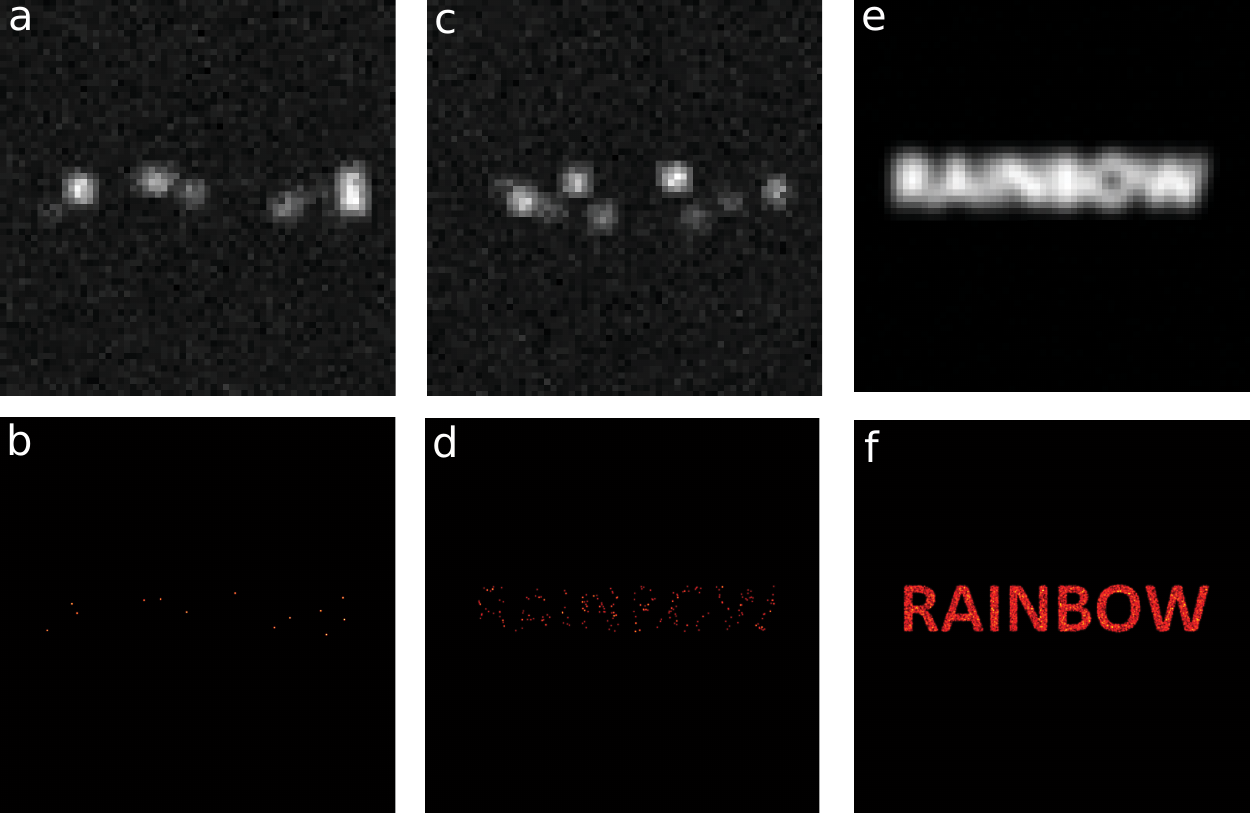}
\caption{SMLM concept. (a, c) Sample frames of raw SMLM  data. (b) Image reconstructed using localizations from a single frame. (d) Image reconstructed using localizations from 50 frames. (e) Diffraction limited image, which is the sum image of 5000 frames of raw SMLM data. (f) Super-resolution image reconstructed using precise localizations from 5000 frames.}
\label{Fig1_5}
\end{figure}

In SMLM, fluorophores can be stochastically switched between fluorescent and dark states, then many images can be taken as a series in each of which only a few fluorophores are emitting. The acquired images are next used to localize isolated fluorophores with much higher precisions than the resolution limit, which is given by
\begin{equation}
\sigma\sim\frac{\sigma_{\mathrm{PSF}}}{\sqrt{I}}
\label{Eq1_6}
\end{equation}
in the absence of background noise, where $\sigma$, $\sigma_\mathrm{PSF}$ and $I$ are the localization precision, size of the PSF and the number of emitted photons by/intensity of the fluorophore, respectively. The resulting precise localizations of these fluorophores can be built up into a super-resolution reconstruction of the underlying structure, Fig. \ref{Fig1_5}. However, major detriments to accurate and precise localizations are: overlapping PSFs when multiple fluorophores in close proximity are activated \cite{wolter2011measuring,small2016multifluorophore}; small photon counts per blinking event resulting in a low signal to noise ratio (SNR); poor labeling of target structures; and low precision in axial location of fluorophores.

Avoiding dense data regions with overlapping PSFs requires sparse activation of fluorophores to prevent activation of more than one fluorophore in a diffraction limited area \cite{rust2006sub,van2011direct,betzig2006imaging,hess2006ultra,lidke2005superresolution,lagerholm2006analysis,egner2007fluorescence}. Some other SMLM approaches overcome the problem of overlapping PSFs by separating the fluorescent signals from molecules with different emission spectra \cite{churchman2005single,bates2007multicolor,low2011erbb1} or lifetimes \cite{thiele2020confocal}. Photo-bleaching of fluorophores has also been employed to achieve low density images of PSFs with minimum overlaps \cite{gordon2004single,qu2004nanometer}. Another promising SMLM technique is based on stochastic binding and unbinding of the diffusing fluorescent emitters to the target, such as DNA-PAINT \cite{schnitzbauer2017super,pallikkuth2018sequential,sharonov2006wide}, and lifeact \cite{riedl2008lifeact,mazloom2021comparing}.

As mentioned, the sparsity constraint imposes activation of a small subset of fluorophores and therefore a few localizations per frame. On the other hand, enough localizations are required to obtain high-resolution images \cite{nieuwenhuizen2013measuring}, which in turn demands undesired long data acquisition times \cite{small2009theoretical,fox2017local}. This problem can be alleviated by multiple-emitter fitting methods that are able to localize emitters in denser regions of the data \cite{small2014fluorophore,sage2015quantitative,sage2019super}; see section 4.3. 

Here, we proceed with discussing the low photon counts and labeling challenges. For various fluorescent probes, the number of emitted photons per blinking event ranges from a few hundred to a few thousand with a small number of blinking events per probe, where more photons and a larger number of blinking events are desired for better localization precision and image contrast \cite{dempsey2011evaluation}.
Furthermore, optimal labeling of the sample is required to achieve high resolution images. For optimal labeling, fluorophores have to only stain target structures with a uniform and high labeling density. Moreover, standard labeling procedures use antibodies to attach fluorophores to a target structure. Some of these antibodies directly bind to the structure and are called primary antibodies. On the other hand, some antibodies bind to target structures via another antibody and are called secondary antibodies. Therefore, the structures that are labeled using secondary antibodies often look artificially larger \cite{zwettler2020molecular}.  

Cellular organelles are inherently 3D structures and therefore 3D microscopy images are desirable. Super-resolution SMLM approaches are able to provide 3D images of cellular structures by exploiting the variance in the PSF shape as a function of axial distance from the focal plane. However, the PSF shape changes slowly along the axial axis and is almost symmetric below and above the focal plane, which hinder precise localization of fluorophres in the axial direction. 
This problem can be overcome by PSF engineering, which intentionally adds aberrations to the microscope to achieve faster changes in PSF and thus higher axial precisions. 3D STORM microscopy was first demonstrated by an engineered astigmatic PSF \cite{huang2008three} and was later perfected using more complex engineered PSFs \cite{pavani2009three,baddeley2011three,shechtman2014optimal,prasad2013rotating} such as the tetrapod PSF \cite{shechtman2014optimal} (see section 4.4). 
\section{Image Formation}
\label{Sec1_4}
Raw SMLM data is comprised of a sequence of image frames, where each frame is a two dimensional array of pixels whose values are a function of the number of photons captured by the camera over a fixed exposure time. Photons reaching the camera often originate from multiple sources of light: 1) the fluorescent light from in-focus fluorophores that label the target structure; 2) the fluorescent light from out-of-focus fluorophores which might be bound to undesired cellular structures, or autofluorescence from these structures; 3) other sources of light that might exist in the area such as cosmic rays. The contribution of light from sources other than the in-focus fluorophores is unwanted and degrades the quality of images \cite{nieuwenhuizen2013measuring,mockl2019bgnet}. The undesired light reaching the camera gives rise to two types of background noise: 1) an approximately uniform, homogeneous background; and 2) a heterogeneous background called structured background \cite{mockl2019bgnet,deschout2014precisely,stetson1987daophot,fazel2019bayesian}.

\begin{figure}[H]
\centering
\includegraphics[scale=1.15]{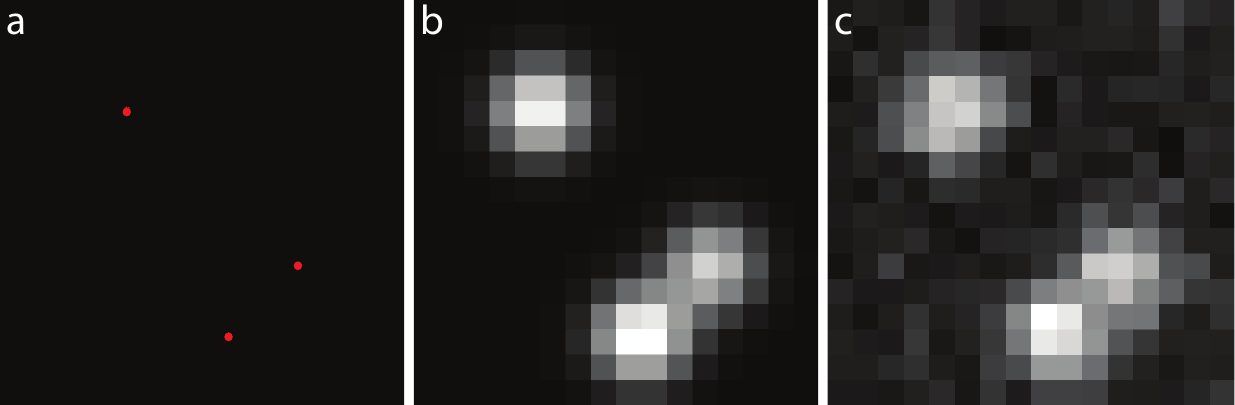}
\caption{Image Formation. (a) Emitter locations. (b) Model is a convolution of emitter locations with the PSF plus a uniform background. (c) The SMLM image recorded by a detector is the model in panel b corrupted with the shot (Poisson) noise. The emitters are assumed to be in-focus.}
\label{Fig1_6}
\end{figure}

For each data frame, a subset of emitters is activated and the image frame is described by 
\begin{equation}
\mathrm{Model}(x,y) = \sum_i I_i \delta(\xi-x_{i})\delta(\eta-y_{i}) \ast \mathrm{PSF}(x,y;\xi,\eta,z_i)+ \mathrm{b},
\label{Eq1_7} 
\end{equation}
which is the convolution of the emitter ${x}$ and ${y}$ locations with the PSF plus a uniform background, Fig. \ref{Fig1_6}. The sum is over the activated emitters. $I_i, x_i, y_i, z_i$ and $\delta$ represent the intensity and location of the $i$th emitter and Dirac delta, respectively. $x$ and $y$ denote image coordinates, and $\xi, \, \eta$ are auxiliary parameters.

Here, we explain different terms present in model (\ref{Eq1_7}) in more detail. First, the additive term in eq. (\ref{Eq1_7}) models the homogeneous uniform background, while the structured background, usually coming from the out-of-focus emitters, is mixed with the in-focus emitters and is given by the convolution term. 
Furthermore, note that while $x$ and $y$ coordinates only determine the lateral location of the PSF, the PSF shape is a function of the ${z}$-location (offset from the focal plane) of the emitter \cite{deschout2014precisely,franke2017photometry}. Therefore, the out-of-focus emitters have a different PSF compared to the in-focus emitters. Moreover, some effects like dipole orientation \cite{engelhardt2010molecular,stallinga2010accuracy}, sample movements \cite{wong2010limit,deschout2012influence} and optical aberrations \cite{stallinga2010accuracy,abraham2009quantitative} result in distortions of the PSF as well. In the absence of these effects, the PSF has an approximate 2D Gaussian form  \cite{wolter2011measuring,abraham2009quantitative,smith2010fast,quan2010ultra,wolter2012rapidstorm,brede2012graspj,starr2012fast,stallinga2010accuracy} 
\begin{equation}
    \mathrm{PSF}(x,y;x_i,y_i,z_i) = \frac{1}{2\pi\sigma_{\mathrm{PSF}}(z_i)^2}\exp\left[-\frac{(x-x_i)^2+(y-y_i)^2}{2\sigma_{\mathrm{PSF}}(z_i)^2}\right].
    \label{GaussPSF}
\end{equation}
However, in cases where the Gaussian PSF is not an appropriate approximation, theoretical PSF models \cite{smith2010fast,mortensen2010optimized,aguet2005maximum} can be used or numerical PSFs can be obtained from calibration experiments \cite{grover2012super,liu2013three,babcock2017analyzing,li2018real}. Such calibrations yield a numeric array of PSF samples which are then utilized to generate a model \cite{liu2013three,li2018real}.

The pixel values recorded by the camera are not the same as the photon counts. Rather, they are stochastic quantities due to noises introduced in the detection process and the stochastic nature of photons. Here, we first explain the main noises added by detection instruments (camera). The camera detectors amplify the signal from the detected photons and a stochastic number of photoelectrons are produced per photon which is used to generate the pixel values \cite{madan1983experimental,heintzmann2016calibrating}. Moreover, the process of converting photoelectrons to pixel values is also a stochastic process that introduces a random offset called the read-out noise, which contributes to the detection noise. To obtain the correct photon counts, therefore, these two effects have to be taken into account.  

We have so far discussed noises introduced by detection instruments, which might be reduced or totally removed by future advances in the instrument technology. However, another major source of stochasticity is shot noise (also called Poisson noise), which comes from the particle nature of photons and cannot be removed regardless of the imaging setup or the used equipment. The expected photon count over a fixed exposure time for every individual pixel can be calculated from model (\ref{Eq1_7}). However, the number of photons captured by a detector over this period slightly deviates from the expected values and has a Poisson distribution, which is called shot noise \cite{ober2004localization,quan2010localization,snyder1993image,smith2010fast}, Fig. \ref{Fig1_6}.

A complete noise model therefore has to consider both detection and shot noises, which might be either pixel dependent (for sCMOS camers \cite{quan2010localization,huang2013video}) or pixel independent (for CCD and EMCCD cameras \cite{madan1983experimental,snyder1993image,heintzmann2016calibrating}). However, such noise models are computationally expensive and thus appropriate approximate forms are employed to formulate image formation for different cameras \cite{snyder1993image,chao2013ultrahigh,quan2010localization,heintzmann2016calibrating,huang2013video,ober2004localization,smith2010fast}.

\section{Image Processing}
\label{Sec1_5}
Image processing is a key step in super-resolution SMLM approaches. This step is comprised of multiple stages: 1) pre-processing; 2) identification of candidate emitters, and localization; 3) filtering and image rendering. Here, we briefly describe each stage and more detailed discussions are postponed to the subsequent sections.

The pre-processing step often alleviates the noise introduced during data acquisition, such as camera and background noise. The estimation of emitter parameters comes next. This step include
finding locations and/or intensities of emitters using the point spread function, which predates the advent of SMLM approaches and had been employed in other scientific disciplines such as electrical engineering \cite{richardson1972bayesian,reddi1979multiple,bobroff1986position}, astronomy \cite{stetson1987daophot,lucy1974iterative,snyder1993image,faisal1995implementation}, and particle tracking \cite{goulian2000tracking,thomann2002automatic}. There are two major classes of localization approaches: single-emitter localization algorithms, and multiple-emitter localization algorithms. The single-emitter algorithms are only able to localize isolated emitters where there are no overlapping PSFs. Single-emitter candidates are usually identified by applying a threshold to detect local maxima and then ROIs of a certain size including those local maxima are selected for further analysis \cite{thompson2002precise,smith2010fast,quan2010ultra,wolter2010real}. Other common detection algorithms employ a wavelet transform \cite{andersson2008localization,kechkar2013real,izeddin2012wavelet}, different types of filters \cite{ma2012fast,wolter2011measuring} and other techniques \cite{smal2008multiple,li2013fast} to identify candidate emitters. The performance of these detection algorithms is highly correlated with SNR and signal to background ratio. For a comparison of different detection algorithms see \cite{smal2009quantitative,ruusuvuori2010evaluation}. After detecting the candidate emitters, single-emitter localization procedures estimate their locations employing either a non-iterative or an iterative algorithm.   

However, a main challenge of the single-emitter algorithms is localizing emitters in dense regions of data with multiple overlapping PSFs from closely spaced activated emitters, Figs. \ref{Fig1_6}-\ref{Fig1_8}. Such dense regions can be generated due to either dense labeling of the sample or fast data collection. Multiple-emitter fitting algorithms are capable of localizing emitters in such dense regions of the data with overlapping PSFs. There are many such multiple-emitter approaches that may be categorized based on their outputs \cite{small2014fluorophore} or based on the algorithm itself \cite{sage2015quantitative,sage2019super}.

After the localization stage, there is often a rejection step that filters out bad localizations to reduce artifacts in the final reconstructions \cite{burgert2015artifacts,reichel2019artifact}.  A popular filtering criteria is based on the found parameter values and their uncertainties that removes the localizations with uncertainties larger than given thresholds \cite{rust2006sub,burgert2015artifacts,fazel2019bayesian}. An additional filtering approach is based on the nearest neighbor distances (NND) where localizations with less than $N$ neighbors within a certain distance are eliminated from the final list of localizations \cite{endesfelder2014simple,heydarian2018template}. 

An example of bad localizations is fitting two or more overlapping PSFs as a single emitter which gives rise to artifacts and degrades image quality. To reduce this effect, the localization algorithms make use of different criteria for recognizing these types of bad fits, Fig. \ref{Fig1_8}. This can be done before localization where ROIs with overlapping PSFs are recognized by a deviation in shape of the identified bright blob from the given PSF \cite{rust2006sub}. Alternatively, these types of bad fits can be recognized after the localization stage by using a $\mathrm{p}$-value test. In this test, the p-values are calculated assuming every individual localization results from a single PSF (the null hypothesis). If the found p-values are smaller than a given threshold, the fits are rejected \cite{smith2010fast}. Finally, the remaining localizations are used to reconstruct a super-resolved image, Fig. \ref{Fig1_8}. 
\subsection{Background Detection}
\label{Sec1_5_1}
The ultimate objective of SMLM techniques is reconstructing high resolution images from precise and accurate \cite{deschout2014precisely} estimates of the emitter locations from raw SMLM data. In order to accomplish this goal, a correct measure of background noise is required as incorrect background leads to biased position estimates \cite{small2014fluorophore}. The correction of uniform background noise is simple and various approaches have been conceived to address this issue. These approaches usually select a ROI and use that to compute a local uniform background noise as the average of pixel values, median of pixel values, average of pixel values after bleaching of the fluorophores, $X$th percentile of pixel values or estimating the additive term in eq. (\ref{Eq1_7}) using an iterative approach \cite{franke2017photometry,smith2010fast,ha1996probing,piccardi2004background,roy2008practical,preus2016optimal}. 

Structured background is significantly more complicated to remove and its presence results in poor position estimates. A few methods have been put forward to cope with this problem including an approach that uses a temporal median filter to subtract structured background from the signal \cite{hoogendoorn2014fidelity}. This technique inspects the fluctuations of pixel values over time to find the background value. The found background can be overestimated in dense regions of the data where there is at least one active emitter at each time. An alternative procedure detects all the emitters regardless of being signal or background and then sets a threshold, removing the emitters with intensities below that as structured background \cite{tang2016automatic}. Wavelet decomposition has also been employed to subtract structured background as well as uniform background prior to emitter detection \cite{min2014falcon}. Recently, a deep learning method has been proposed to detect structured background using the PSF shape \cite{mockl2019bgnet}. Moreover, Faze1, et al. used a Bayesian approach to model structured background with a collection of PSF-sized dim emitters \cite{fazel2019bayesian}. In the field of astronomy, methods such as sigma clipping has been developed to deal with structured background in dense data sets \cite{stetson1987daophot}. In the sigma clipping procedure, the brightness mean, $m_{\mathrm{I}}$, and standard deviation, $\sigma_{\mathrm{I}}$, are calculated and those intensities outside the range of $[m_{\mathrm{I}}-\alpha \sigma_{\mathrm{I}},\, m_{\mathrm{I}}+\alpha \sigma_{\mathrm{I}}]$ are considered noise \cite{stetson1987daophot}.
\subsection{Single Emitter Fitting}
\label{Sec1_5_2}
The single-emitter localization algorithms can be classified into two major categories: the algorithms that use non-iterative approaches to localize emitters and the algorithms that use an iterative procedure. Studies show that the iterative algorithms are more accurate than the non-iterative algorithms \cite{cheezum2001quantitative}. However, iterative algorithms are computationally more demanding and require a precise PSF model.
\subsubsection{Non-iterative Algorithms}
Non-iterative algorithms do not need any information about the PSF and are usually fast and easy to implement. However, they are not as accurate as iterative algorithms that utilize the PSF to generate a model of the data. The lack of enough accuracy is often a consequence of different types of noise.     

A few non-iterative approaches such as QuickPALM \cite{henriques2010quickpalm} calculate the emitter locations as center of mass of ROIs containing single emitters \cite{henriques2010quickpalm,pavani2009three}. This gives a good estimate of location, however, failure in background correction results in biased localizations towards the center of the ROIs. Virtual Window Center of Mass (VWCM) \cite{berglund2008fast} ameliorates this issue by iteratively adjusting the selected ROI to minimize the separation of the emitter location and the center of the ROI.

FluoroBancroft borrows the Bancroft procedure from the satellite Global Positioning System (GPS) to localize emitters \cite{andersson2008localization,hedde2009online}. This approach uses three pixel values within a PSF to draw three circles. The emitter is located within the intersection of these three circles. The size of the intersection region is a measure of the localization precision. A correct measure of background is also of great importance in this approach to calculate accurate radii.

Single emitters can also be localized by finding the gradient of the phase of the Fourier transform of the ROIs. For a single emitter, eq. (\ref{Eq1_7}) reduces to
\begin{equation}
I(m,n)=I_0\delta(\xi-x_0)\delta(\eta-y_0) \ast \mathrm{PSF}(x_m,y_n;\xi,\eta,z_0)+b
\label{Eq1_8}
\end{equation}
where $m$ and $n$ count rows and columns of pixels in the ROI, and $x_m$ and $y_m$ give the centers of those pixels. $(x_0,y_0,z_0)$ is the location of the single emitter. The Fourier transform of eq. (\ref{Eq1_8}) produces
\begin{equation}
\tilde{I}(k,l)=H(k,l) \, \mathrm{exp}\left[-i2\pi \left(\frac{x_0}{M}k+\frac{y_0}{N}l\right) \right]+\tilde{b}
\label{Eq1_9}
\end{equation}
where $k, l$ and $M, N$, respectively, denote pixel row and column indices and the maximum number of pixels in rows and columns. Furthermore, $H$ is a real quantity. For data sets with large SNR, the background term is negligible and the Fourier Domain Localization Algorithm (FDLA) gives the emitter position as the average of the gradient of the phase \cite{yu2011fast}:
\begin{equation}
x_0=\mathrm{mean}\left(\left(\frac{\partial \phi}{\partial k}\right)\frac{M}{2\pi}\right), \,\,\,\, y_0=\mathrm{mean}\left(\left(\frac{\partial \phi}{\partial l}\right)\frac{N}{2\pi}\right)
\label{Eq1_10}
\end{equation} 
where $\phi = \mathrm{arctan}\frac{\mathrm{Im}(\tilde{I})}{\mathrm{Re}(\tilde{I})}$. The performance of this approach suffers from the presence of background noise as well. Another approach localizes single emitters by calculating the first Fourier coefficients in both $x$ and $y$ directions, and the phase of these coefficients are then employed to find the emitter location \cite{martens2018phasor}.  

Radial symmetry of the PSF has also been employed to calculate emitter locations \cite{ma2012fast,parthasarathy2012rapid}. Due to the radial symmetry of PSFs, intensity gradient vectors for different pixels converge to the region with maximum intensity where the emitter is located. This approach is robust in the presence of uniform background noise and achieves precisions close to the theoretical lower limit of precision given by the Cramer-Rao Lower Bound (CRLB).
\subsubsection{Iterative Algorithms}
Iterative algorithms are the most rigorous approaches for emitter localization. In these approaches, the parameters in model (\ref{Eq1_8}) are adjusted in an iterative manner to fulfill a certain criterion. In the localization problem, the parameters are ($x_0, y_0, z_0, I_0, b$), the emitter location, the number of photons from (intensity of) the emitter and a uniform background noise. The criteria that are extensively utilized in the emitter fitting literature are the Least Square (LS) difference between data and model, and maximizing the likelihood function via a Maximum Likelihood Estimate (MLE). 

CRLB states that the fundamental limit of variance for estimating a parameter from given data is obtained by the inverse of the Fisher Information \cite{thompson2002precise,ober2004localization,ram2006beyond}. Therefore, the fundamental limit of precision is given by the inverse of the square root of the Fisher Information. Theoretically, MLE achieves the best localization precision, equivalent to the CRLB \cite{ober2004localization,deschout2014precisely,smith2010fast,abraham2009quantitative,ram2006beyond,huang2011simultaneous,mortensen2010optimized,rieger2014lateral,baddeley2018biological}. LS performance is comparable to the MLE under certain conditions described below \cite{abraham2009quantitative,mortensen2010optimized,rieger2014lateral,baddeley2018biological,von2017three}. 

The performance of weighted LS approaches that of MLE at high SNR, when the Poisson noise (shot noise) can be well approximated by a Gaussian model, or when read-out noise is dominant. Note that neither of these scenarios are correct for SMLM data where the read-out noise is usually negligible in the presence of Poisson noise and the SNR is not too high. In general, MLE yields better localization accuracy and is more robust in the presence of PSF mismatch, but is computationally more complex and requires an accurate model of noise \cite{abraham2009quantitative,mortensen2010optimized}. \\

\noindent
\textbf{Least Squares Fitting.}
The LS approaches iteratively vary the parameters of the model to minimize the sum of differences between the pixel values from the data and the model. The developed algorithms use a Gaussian PSF \cite{thompson2002precise,abraham2009quantitative}, theoretical PSFs \cite{kirshner20133,ovesny2014thunderstorm} or experimentally acquired PSFs \cite{mlodzianoski2009experimental,kirshner2013can} to calculate a model using eq. (\ref{Eq1_8}). The difference of the generated model and data is then given by
\begin{equation}
D=\sum_{\mathrm{pixel}}\, \frac{\left(\mathrm{data}-\mathrm{model}\right)^2}{\mathrm{expected\,variance}}
\label{DiffLS}
\end{equation}
where in weighted LS (WLS) the differences are scaled by the expected variance of the noise, which scales the errors for individual pixels \cite{pavani2009three,abraham2009quantitative,mlodzianoski2009experimental,wolter2012rapidstorm,kirshner20133,ovesny2014thunderstorm}. A pixel with a high signal is expected to have a large noise variance, and therefore it is allowed to have a larger error in the WLS procedure. However, the scaling factor is replaced by one in the unweighted LS algorithm, which we call LS hereafter, and does not accommodate noises \cite{thompson2002precise,wolter2010real,juette2008three,kirshner2013can,zhu2013efficient}. 
The Levenberg-Marquardt iterative procedure \cite{kirshner20133,kirshner2013can,zhu2013efficient}, or other procedures \cite{mlodzianoski2009experimental}, are then employed to iteratively adjust the parameters of the model to reduce the difference (\ref{DiffLS}). 

The WLS algorithm accomplishes accuracies close to the CRLB when the photon count is high, but the noise variance needs to be known as well as an accurate PSF model. The PSF mismatch, particularly in the tail of the PSF, results in large errors when scaled by a small expected noise variance in the pixels far from the emitter \cite{small2014fluorophore}.  Therefore, the LS algorithm is more suitable when a reasonable PSF model and/or noise model are not accessible. \\

\noindent
\textbf{Maximum Likelihood Estimator.}
To calculate the likelihood, we first need to find expected photon counts per pixel. The contribution of an emitter located at $(x_0,y_0,z_0)$ to the $k$th pixel's photon count is given by
\begin{equation}
\Delta_k = \frac{I_0}{2\pi \sigma_{\mathrm{PSF}}(z_0)^2}\int_{x_k-0.5}^{x_k+0.5}\int_{y_k-0.5}^{y_k+0.5} \mathrm{exp}\left[\frac{(x-x_0)^2+(y-y_0)^2}{2\sigma_{\mathrm{PSF}}(z_0)^2}\right]dx\,dy
\label{Eq1_12}
\end{equation}
where we used the Gaussian PSF (\ref{GaussPSF}). $\Delta_k, \sigma_{\mathrm{PSF}},I_0, x_k$ and $y_k$, respectively, denote the number of photons in the $k$th pixel from the emitter, the half width of the Gaussian PSF, total number of photons from the emitter, the emitter location and the center of the $k$th pixel.

\begin{figure}[H]
\centering
\includegraphics[scale=0.82]{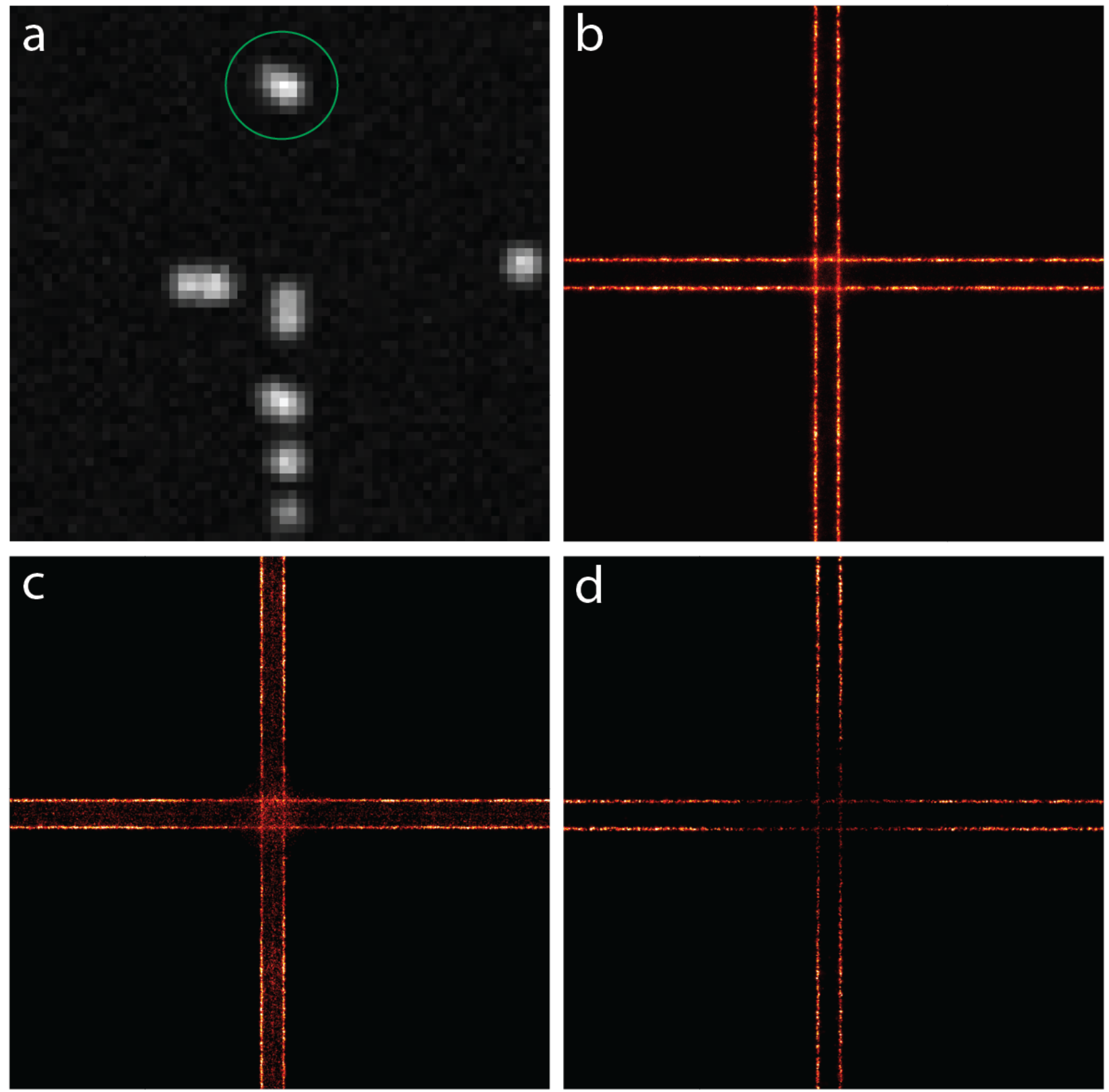}
\caption{Reconstructions from dense data with overlapping PSFs. (a) A frame of dense raw SMLM data of a cross with parallel lines. The green circle shows an example of two overlapping PSFs. (b) Reconstruction from a multiple-emitter algorithm with no filtering. (c) Reconstruction from single-emitter algorithm with no filtering. The localizations in the area between the lines are the results of fitting overlapping PSFs with a single PSF. (d) Reconstruction from single-emitter algorithm after filtering. The dense regions of data appear sparse when processed with a single-emitter algorithm due to the inability to localize overlapping PSFs, which is called the contrast inversion artifact.}
\label{Fig1_8}
\end{figure}

The total photon count in the $k$th pixel is the sum of the photons from the emitter and the uniform background noise
\begin{equation}
\lambda_k=\Delta_k+b.
\label{Eq1_13}
\end{equation}
Equation (\ref{Eq1_13}) yields the expected number of photons for pixel $k$ for a fixed exposure time. Consequently, the number of detected photons in pixel $k$ has a Poisson distribution, which gives the likelihood of the $k$th pixel 
\begin{equation}
P_k(D|\theta)=\frac{\lambda_k^{D_k}e^{-\lambda_k}}{D_k!}
\label{Eq1_14}
\end{equation}
where $\theta$ stands for the set of parameters ($\theta=(x_0,y_0,z_0,I_0,b)$, where $z_0=0$ for 2D cases). $D$ represents data, which is a two dimensional array of pixels whose values are related to the number of photons detected by the camera. $D_k$ selects the $k$th pixel in $D$. Since the pixels are independent, the likelihood of the ROI can be written as the product of the likelihoods of all the pixels in the ROI
\begin{equation}
P(D|\theta)=\prod_k P_k(D|\theta).
\label{Eq1_15}
\end{equation}

The two main iterative algorithms employed in the literature to find the parameters that optimize the above likelihood are variations of the Newton method \cite{abraham2009quantitative,smith2010fast,brede2012graspj,starr2012fast,aguet2005maximum,liu2013three,babcock2017analyzing} and a modified version of Levenberg-Marquardt \cite{wolter2011measuring,wolter2012rapidstorm,li2018real,laurence2010efficient} adopted from LS procedures.  

The Newton approach is employed to find the root of the derivative of the likelihood function (\ref{Eq1_15}), and therefore one needs to calculate the second derivative of the likelihood as well, which is computationally demanding. On the other hand, the Levenberg-Marquardt algorithm only calculates the first derivative of the likelihood, which makes it computationally less demanding in comparison \cite{wolter2012rapidstorm,li2018real}. Different strategies have been exploited to speed up the Newton optimization algorithm including implementation on Graphical Processing Units (GPUs), which allows parallel analysis of ROIs \cite{smith2010fast,quan2010ultra,brede2012graspj,starr2012fast,babcock2017analyzing}; starting from better initial values \cite{mortensen2010optimized}; and estimating $x$ and $y$ positions individually utilizing the separability property of the Gaussian function \cite{starr2012fast}. 

\subsection{Multiple Emitter Fitting}
\label{Sec1_5_3}
Raw SMLM data is often acquired via a wide-field illumination procedure where the whole sample is exposed to the  excitation laser. Emitter activation is a stochastic process, and hence there will always be activated emitters at close proximity. Therefore, overlapping PSFs are unavoidable, even under sparse activation conditions. The overlapping PSFs are eliminated in a filtering step in single-emitter approaches, which results in losing information \cite{huang2011simultaneous} as well as the appearance of artifacts, for instance, contrast inversion, Fig. \ref{Fig1_8}.   

The inability of single-emitter algorithms to localize multiple activated emitters in a diffraction limited vicinity enforces the sparse activation of emitters. This is followed by a long acquisition time to build a full list of localizations to reconstruct an image with high resolution. In some experiments, such as studies of live or dynamic samples, fast data acquisition is preferred and hence dense activation of emitters is inevitable. 
Therefore, proper analysis of dense SMLM data with overlapping PSFs is necessary to reduce data acquisition time, avoid artifacts and facilitate live sample imaging. 

Numerous multiple-emitter fitting algorithms have been devised to fit emitters with overlapping PSFs, some borrowed from other areas, such as astronomy and statistical analysis. The reported algorithms employ a very wide range of approaches and have a broad spectrum of performance \cite{sage2015quantitative,sage2019super}. These procedures are often iterative algorithms or include an iterative training step.
\subsubsection{Least Squares}
The LS algorithm has been employed to fit multiple-emitters in the field of astronomy \cite{stetson1987daophot}. It was modified for SMLM super-resolution microscopy, called DAOSTORM in this context \cite{holden2011daostorm}. DAOSTORM uses isolated emitters in the raw data to estimate the PSF, and then uses the found PSF to fit emitters in dense data sets. The algorithm starts with an initial number of emitters located at the brightest pixels of the ROI, and then uses least squares fitting to localize them with sub-diffraction precision. Next, the residuum image is calculated by subtracting the model from the data, and is used to detect new emitters in the pixels brighter than a given threshold. The detected emitters are then localized to obtain sub-diffraction precision. This step is repeated until there is no pixel with intensity above the threshold in the residuum image.  
\subsubsection{Maximum Likelihood}
The MLE approach that was described before can be modified for multiple-emitter fitting within ROIs with overlapping PSFs. The total photon counts in the $k$th pixel is given by
\begin{equation}
\lambda_k(N)=b+\sum_{i=1}^N \Delta_{k,i}
\label{Eq1_16}
\end{equation}
where $\Delta_{k,i}$ is the number of photons received in the $k$th pixel from the $i$th emitter and can be calculated using (\ref{Eq1_12}). $N$ and $b$ are the number of emitters and the uniform background. The likelihood of the $k$th pixel is then
\begin{equation}
P_k(D|\theta)=\frac{\lambda_k(N)^{D_k}e^{-\lambda_k(N)}}{D_k!}.
\label{Eq1_17}
\end{equation}
The likelihood of the ROI is obtained from the product of the likelihoods of individual pixels (\ref{Eq1_15}). 

The likelihood (\ref{Eq1_17}) has more than one emitter and therefore more parameters to estimate, demanding more iterations and computational time. The MLE approach is implemented in the same manner as single-emitter fitting to estimate the parameters. Nevertheless, there is a new parameter, $N$, the number of emitters, which cannot be directly estimated from the likelihood itself. This is because adding more emitters tends to increase the likelihood, known as over-fitting. The approaches that find the number of emitters are called model selection algorithms. Several model selection algorithms have been reported along with the MLE localization procedure, including thresholding of the residuum image \cite{babcock2012high}, p-value of the likelihood ratios \cite{huang2011simultaneous}, Bayesian Information Criteria (BIC) \cite{quan2011high,wang2012palmer}, PSF size and NND \cite{li2019divide}, and others \cite{ashida2016precise,sun2014superresolution,aristov2018zola}.

The 3D-DAOSTORM \cite{babcock2012high} is a 3D multiple emitter fitting approach. 3D fitting procedures will be discussed in the next section and here we explain how this procedure deals with overlapping PSFs. 3D-DAOSTORM fits overlapping PSFs by fitting only choices from the brightest emitters in the ROI at the beginning. It then subtracts the obtained model from the ROI intensities and uses the residuum image to find pixels brighter than a given threshold to detect new emitters. It employs MLE to localize the new emitters. The new emitters are added to the list of detected emitters and this step is repeated until there is no pixel brighter than the given threshold.

Simultaneous multiple-emitter fitting \cite{huang2011simultaneous} starts from a model with one emitter, $N=1$, and goes up to $N=N_{max}$. For each model, this method localizes the emitters employing the MLE approach. The log-likelihood ratio (LLR)
\begin{equation}
\mathrm{LLR}=-2\log\left[\frac{P(D|\hat{\theta})}{P(D|D)}\right]
\label{Eq1_18}
\end{equation}
has an approximate chi-square distribution \cite{huang2011simultaneous}, where $\hat{\theta}$ is the parameters that maximize the likelihood and $P(D|D)$ gives the upper limit of the likelihood. This LLR is then used to calculate p-values where the model with lowest $N$ that meets a threshold p-value is accepted as the best fit. 

As mentioned above, the MLE approach suffers from over-fitting, and adding more parameters (emitters) tends to give larger values for likelihoods. Bayesian Information Criteria is a model selection algorithm that penalizes adding new parameters to an MLE problem. SSM-BIC \cite{quan2011high} selects the model that maximizes the following function
\begin{equation}
\mathrm{BIC}(N)=\frac{(\mathrm{Data}-\mathrm{Model})^2}{\mathrm{Data}}+(3N+1)\log{(m)}
\label{Eq1_19}
\end{equation}
where $m$ is the number of pixels in the given ROI. Note that there are $3N+1$ parameters in a 2D model with $N$ emitters and a known PSF. This approach has also been implemented on GPUs with $\sim$100 times decrease in computational time \cite{wang2012palmer}. 

QC-STORM \cite{li2019divide} uses a weighted likelihood
\begin{equation}
P_W(D|\theta)=\prod_k W_k P_k(D|\theta)
\label{Eq1_20}
\end{equation}
to localize emitters, where $W_k$ is the weight of the $k$th pixel and is smaller for pixels closer to the edges of the ROI. The weighted likelihood therefore suppresses the signal close to the edges of the ROI which makes it an effective method to localize emitters within ROIs with signal contaminations close to their edges. QC-STORM identifies ROIs with more than one emitter based on the ROIs' NNDs and the size of the PSF estimated using weighted MLE. This algorithm has been implemented on GPUs and is capable of processing very large fields of view.

Some other approaches use an iterative deconvolution algorithm to accomplish maximum likelihood employing the Richardson and Lucy procedure \cite{mukamel2012statistical,zhao2018faster}. These approaches return a grid image with finer pixel sizes than the camera pixel size, with non-zero pixel values at the emitter locations rather than returning a list of localizations. 
\subsubsection{Bayesian Inference}
The MLE algorithm suffers from overfitting as discussed above. Bayes' formula (\ref{Eq1_21}) provides an elegant way to include prior knowledge into the problem, allowing the model to be restricted to reasonable number of parameters. It however adds complications to the problem by including more distributions. Furthermore, it has been shown that the Bayesian approach can achieve localization uncertainties better than those from MLE by inclusion of reasonable prior knowledge \cite{linden2017pointwise}. The Bayesian paradigm is equivalent to MLE when there is no prior knowledge available. The posterior is given by
\begin{equation}
P(\theta|D)=\frac{P(D|\theta)P(\theta)}{P(D)}
\label{Eq1_21}
\end{equation}
where $P(\theta)$ and $P(D)$ are, respectively, the prior on the parameters $\theta$ and the normalization constant, 
\begin{equation}
P(D)=\int P(D|\theta)P(\theta) d\theta
\label{Eq1_22}
\end{equation}  
called the evidence. Another difference of MLE and Bayesian approaches is that MLE returns fixed emitter parameters (location and intensity), while the Bayesian procedure returns a probability distribution, the posterior, for every emitter parameter. 
A few fully Bayesian algorithms have been developed for multiple-emitter fitting so far \cite{fazel2019bayesian,cox2012bayesian,li2020live}, where we discuss two of them here. 

The Bayesian multiple-emitter fitting (BAMF) algorithm \cite{fazel2019bayesian} employs the Reversible Jump Markov Chain Monte Carlo (RJMCMC) \cite{green1995reversible,richardson1997bayesian} technique to explore model spaces with different dimensions, or equivalently different number of emitters, and returns a posterior distribution which is a weighted average of different possible models. BAMF uses either a Gaussian PSF model or an arbitrary input numerical PSF along with an empirical prior on emitter intensities to attain precise and accurate localizations in dense regions of data. This technique 
performs emitter fitting, model selection and structured background modeling simultaneously, and therefore takes into account various sources of uncertainties which are often ignored.  

The 3B algorithm \cite{cox2012bayesian} analyzes the entire data set at the same time by integrating over all possible positions and blinking events of emitters. 
This algorithm makes inferences about the emitter locations, intensities, width of the Gaussian PSF and blinking of emitters. The posterior of this problem is given by
\begin{equation}
P(a,b,N|D)=\frac{P(D|a,b,N)P(a)}{P(D)}
\label{Eq1_23}
\end{equation}
where $a, b$ and $N$ represent the emitter parameters, blinking events and the number of emitters, in turn. 3B uses a uniform prior for the locations and a log-normal prior for other parameters. The discrete parameter, $b$, is then integrated out using an MCMC \cite{hastings1970monte} approach to obtain the posterior distribution of $a$, $P(a,N|D)$. Next, the Maximum A Posteriori (MAP) is computed using the conjugate gradient approach to obtain the emitter locations, intensities and the PSF size. After that, the parameter $a$ is also marginalized to get the model probability, $P(N|D)$, for model selection. 

The 3B algorithm models the entire data set and is able to use all the collected photons from multiple blinking events to achieve better localization precision. The returned image is a probability map of the weighted average of all possible models rather than a selected single model \cite{lidke2012super}. Moreover, it needs a simple experimental setup for data collection \cite{cox2012bayesian}. However, this technique is very slow because calculating the integrals to marginalize parameters $a$ and $b$ is extremely computationally demanding. There have been several attempts to speed up the algorithm including 3B implementation in cloud computing \cite{hu2013accelerating}, use of more informative priors \cite{xu2015bayesian}, and initializing the algorithm with better starting parameter values \cite{xu2017live}. 
\subsubsection{Compressed Sensing}
A frame of SMLM data can be considered as a matrix $y$
\begin{equation}
y=Ax+b
\label{Eq1_24}
\end{equation}
where $x$ is the signal, which is an up-sampled discrete grid (image) with non-zero elements at the emitter locations, $A$ is the PSF matrix, and $b$ is the uniform background. The objective is to recover the non-zero elements of the signal $x$ where most of the elements are zero due to sparse activation of fluorescent emitters in SMLM microscopy. Compressed Sensing (CS) theory states that a signal $x$ can be recovered from a noisy measurement $y$ if the signal is sufficiently sparse \cite{candes2006stable}. This mathematically can be expressed as
\begin{align}
\mathrm{Minimize\hspace{-1mm}:} &\,||x||_1 \hspace{2.8cm}\\
\mathrm{Subject\,to\hspace{-1mm}:} &\,||y-(Ax+b)||_2 \leq \epsilon 
\label{Eq1_25}
\end{align}
where $||x||_1=\sum_i|x_i|$ is the L1-norm of the up-sampled image, and the L2-norm of the residuum image is given by
\begin{equation}
    ||y-(Ax+b)||_2=\sqrt{\sum_i\left(y_i-(Ax+b)_i\right)^2}.
\end{equation}
The inequality allows fluctuations from the expected model due to different types of noise. 

Various optimization approaches have been utilized to minimize the L1-norm in the presence of the given restriction in (\ref{Eq1_25}) including a convex optimization algorithm \cite{zhu2012faster}, L1-Hotomopy \cite{babcock2013fast}, gradient descent \cite{min2014falcon}, and others \cite{hugelier2016sparse,gazagnes2017high}. These algorithms are able to detect and localize the emitters in very dense regions of the data. However, due to the large size of the up-sampled image, the CS algorithms are slow and the resolution cannot be better than the grid size of the up-sampled image.     

The issues mentioned above have been addressed in later literature using different approaches. FALCON \cite{min2014falcon} accelerates the algorithm by implementing CS on GPUs. It also ameliorates the grid-size problem by refining the found locations in the subsequent steps after the deconvolution stage. CS has recently been implemented over continuous parameter spaces to remove the limits imposed by the up-sampled grid \cite{boyd2017alternating,huang2017super}. To lower the computational cost of the CS algorithm, a recent paper models the entire sequence at the same time rather than employing a frame by frame analysis of the data \cite{wu2018fast}. Another approach implements the CS algorithm in the correlation domain by calculating the frame cross-correlations \cite{solomon2018sparsity}.    
\subsubsection*{Singular Value Decomposition}
Assuming $A$ as an $n\times n$ square matrix with non-zero determinant, it can be factorized into
\begin{equation}
A=U\Sigma U^{-1}
\label{Eq1_Sing}
\end{equation}
where U is an $n\times n$ matrix whose columns are the eigenvectors of the decomposition, and $\Sigma$ is a diagonal $n \times n$ matrix where the diagonal elements are the eigenvalues. A non-square matrix $B$ can also be decomposed in a similar fashion, called the singular value decomposition (SVD)
\begin{equation}
B_{n\times m}=V_{n\times n}\Lambda_{n\times m} W_{m\times m}
\label{Eq1_27}
\end{equation}
where $V$ and $W$ are, respectively, $n\times n$ and $m\times m$ matrices. $\Lambda$ is a diagonal $n\times m$ matrix with diagonal elements the square roots of the eigenvalues of $BB^T$, where $T$ stands for transpose \cite{nielsen2002quantum}. 

The MUltiple SIgnal Classification ALgorithm (MUSICAL) for super-resolution fluorescence microscopy \cite{agarwal2016multiple} takes $B$ as a collection of frames of super-resolution images where each column of $B$ is a frame of the raw data. $B$ is then a non-square matrix that can be factorized into a diagonal matrix and unitary matrices where the eigenvectors are eigenimages. The eigenimages are next classified into signal and noise based on the corresponding eigenvalues using a given threshold. Finally, MUSICAL calculates the projection of the PSF at different locations in the eigenimages to identify and localize the emitters. An alternative method makes use of the SVD in the Fourier domain, analyzing the sequence of raw data frame by frame to localize the emitters \cite{huang2015fast}. 
\subsubsection*{Deep Learning}
Deep learning approaches are non-iterative optimization algorithms that perform calculations in a parallel manner and hence are very fast. Theses approaches are based on
Artificial Neural Networks (ANN) that are inspired by animal brains and neural systems. The building blocks of the brain are neurons equivalent to perceptrons or sigmoid neurons in ANNs \cite{nielsen2015neural}, which are explained in the following. 

\begin{figure}[H]
\centering
\vspace{-12mm}
\includegraphics[scale=1.4]{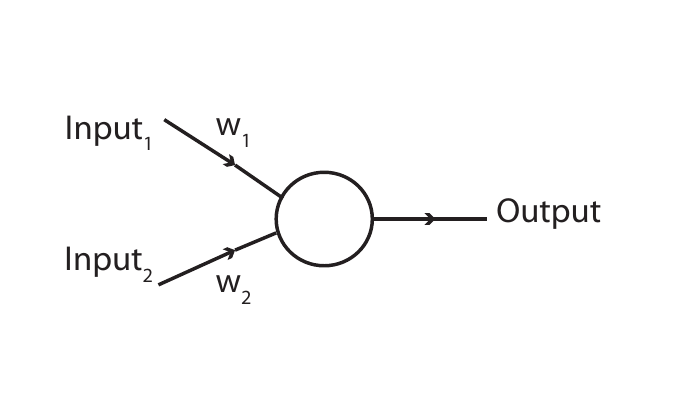}
\vspace{-8mm}
\caption{Perceptron takes several binary inputs and gives a binary outcome. $w_i$s are the weights of the inputs.}
\label{Fig1_9}
\end{figure}

Perceptrons take a few binary inputs and generate a binary output, Fig. \ref{Fig1_9}.
The output of the perceptron is one if the sum of inputs times their weights is larger than a threshold and is zero otherwise, eq. (\ref{Eq1_26}).

\begin{equation}
\mathrm{Output}=\begin{cases}
0 \hspace{0.5cm} \mathrm{if} \hspace{0.5cm} \sum_i w_i \,\, \mathrm{input}_i < \mathrm{threshold}\\
1 \hspace{0.5cm} \mathrm{if} \hspace{0.5cm} \sum_i w_i \,\, \mathrm{input}_i > \mathrm{threshold}
\end{cases}
\label{Eq1_26}
\end{equation}\\
It can be shown that certain combinations of perceptrons produce logical operations such as AND, OR and NAND, which are the underlying bases of computation, and any mathematical function could be generated using them \cite{nielsen2015neural}. Therefore, ANNs are capable of producing any mathematical function using perceptrons. A network of perceptrons can be trained to perform different tasks by adjusting the weights, $w_i$. On the other hand, sigmoid neurons are more sophisticated versions of perceptrons where the output is a value within the interval of $[0, 1]$ rather than a binary output. Sigmoid neurons are more flexible in the training process and are used in neural networks.  

ANNs have been employed to attack various problems in the field of biomedical imaging \cite{belthangady2019applications,mockl2020deep}, specifically for SMLM image processing \cite{strack2018deep}.
ANNs designed to localize emitters in raw SMLM data can be categorized into three different types based on their training approach:\\
1) ANNs are trained using simulated data where the ground truth is available or using localizations found by a standard iterative algorithm \cite{nehme2018deep,nehme2019dense,zelger2018three}. In the training stage, the ANN learns by minimizing the sum of distances between the found localizations and the given locations. Using synthesized data, there will always be adequate training data.  \\
2) ANNs are also trained using two well-registered sets of data acquired from the same sample where one of them is used as ground truth. The ground truth image has high SNR that can be acquired employing different procedures such as confocal microscopy \cite{weigert2018content,wang2018deep}, using an objective with high numerical aperture \cite{wang2019deep}, or using a sparse data set to reconstruct a super-resolved image with high SNR \cite{ouyang2018deep} to train the network. In the training stage, the network learns by minimizing the difference between the output image and the acquired image with high SNR. The minimization of differences can be implemented via a standard iterative optimization algorithm \cite{weigert2018content}, or by using a sub-network in the training stage, which takes the output of the ANN along with the ground truth and labels the output as real or fake \cite{wang2019deep,ouyang2018deep}.\\
3) In an alternative training approach, there is no ground truth used in the training step. Instead, the network is trained by reproducing the input data from the list of found emitters and minimizing the difference between the original input and the synthesized image \cite{speiser2019teaching}. This training procedure is called unsupervised learning.

The deep learning procedures have multiple advantages including: fast data analysis, no required input parameters or thresholds, and performance comparable to the MLE algorithms \cite{zelger2018three,kim2019information}. However, the training process is very sensitive and has to be done very carefully. Two major pitfalls of training are the hallucination and generalization problems \cite{belthangady2019applications}. Deep learning algorithms might make mistakes in identifying patterns from random inputs when there is not adequate training, which is called the hallucination problem. If there are new patterns that are not seen by the algorithm before, it fits these new data by the old patterns, this is called the generalization problem. 
\subsubsection*{Others}
WindStorm \cite{ma2019windstorm} uses a temporal filter to estimate the background and remove it from the raw data. It then implements deconvolution by dividing the Fourier transform of the clean image with the Fourier transform of the PSF. It next performs frequency truncation. The recovered locations are given by the peaks of the deconvolved image in the spatial domain. This is a fast and non-iterative approach and the found locations can be used as initial values for iterative algorithms.

Wedge Template Matching (WTM) \cite{takeshima2018multi} identifies and localizes emitters by matching an entire or partial template of the PSF to the regions of the data with overlapping emitters. The WTM algorithm picks either an entire or partial PSF template based on the degree of overlapping between PSFs. It finds the candidate pixels containing emitters using cross correlation of the template with the image, and then finds the locations of the detected emitters with sub-pixel accuracies.

Other approaches employ machine learning algorithms \cite{colabrese2018machine}, independent component analysis and a shape matching approach in the frequency domain \cite{barsic2013super}, and other algorithms \cite{simonson2011single} to identify and localize emitters in dense regions of the data.
\subsection{3D Emitter Fitting}
\label{Sec1_5_4}
\begin{figure}[H]
\centering
\includegraphics[scale=0.74]{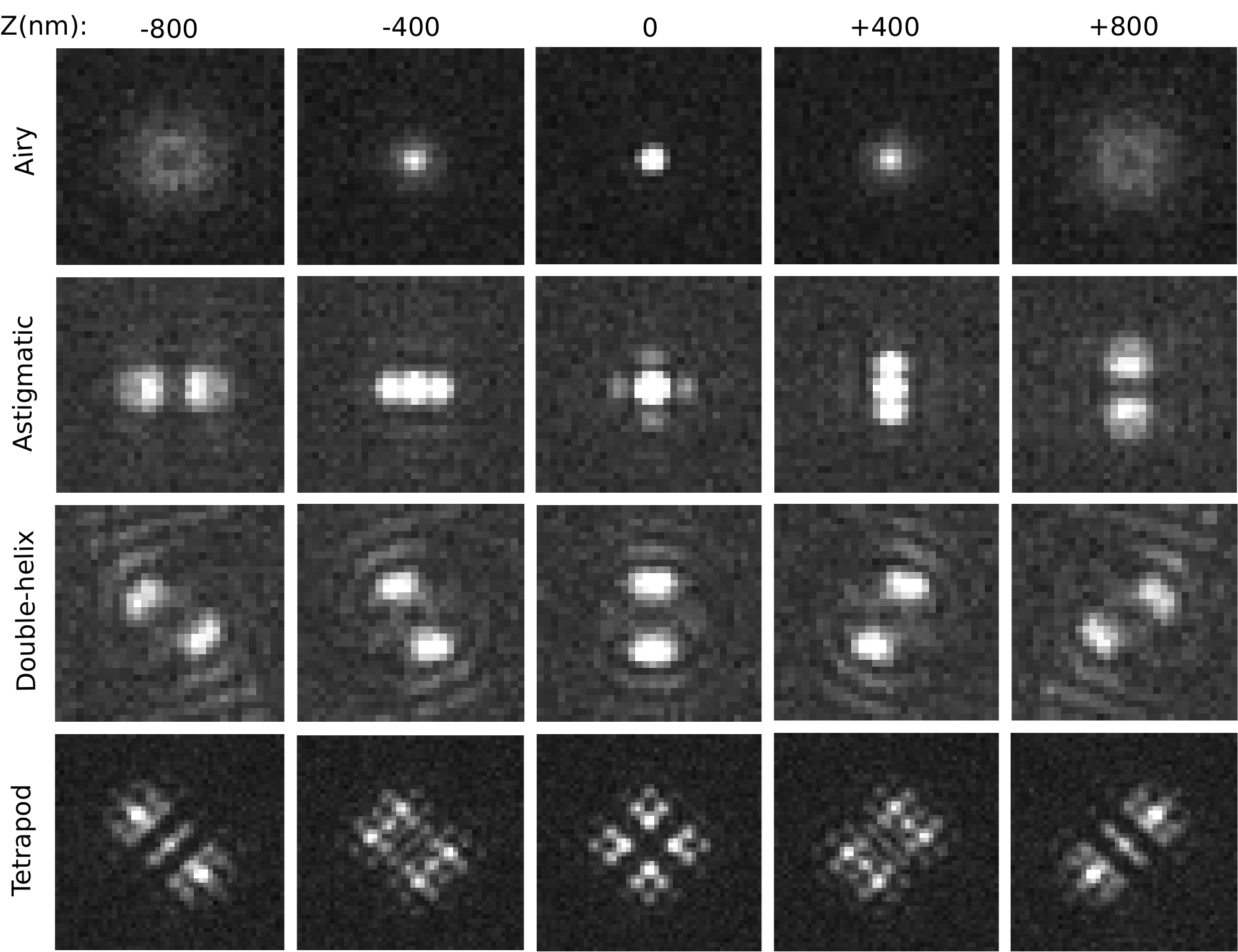}
\caption{Different engineered PSFs at multiple $z$-locations (corrupted with Poisson noise and a uniform background). The Airy pattern has a symmetric shape above and below the focal plane. The astigmatic, double-helix and tetrapod PSFs are engineered to break the symmetry below and above the focal plane and yield better localization precision along the axial direction.} 
\label{Fig_PSF}
\end{figure}

Biological samples are 3D in nature and 3D microscopy approaches are required to gain better insight into biological structures. A standard super-resolution microscope cannot provide precise axial location of an emitter due to slow changes in the 3D PSF as a function of axial position ($z$), and the symmetry of the PSF below and above the focal plane \cite{von2017three}, see the Airy PSF in Fig. \ref{Fig_PSF}. Super-resolution microscopes can however provide precise axial location of an emitter by some alterations in the optical setup. Several different modifications have been reported in the literature encompassing multi-focal methods \cite{mlodzianoski2009experimental,prabhat2004simultaneous,ram2008high,tahmasbi2014designing} which image multiple focal planes at the same time; engineered PSFs \cite{huang2008three,pavani2009three,baddeley2011three,shechtman2014optimal,prasad2013rotating,lew2011corkscrew,jia2014isotropic} where the axial information is encoded into the PSF shape; and other procedures \cite{moerner1994near,ruckstuhl2004supercritical,jung2016three,fu2016axial}. Here, we briefly discuss the multi-focal and PSF engineering techniques.       

The multi-focal approaches allow precise axial emitter localization by providing multiple $z$-slices of the PSF. This can be achieved by splitting the emission beam into multiple light-paths with different optical path lengths \cite{prabhat2004simultaneous,ram2008high,liu2013three}. The lateral locations of the PSF are often registered to be the same on all paths and the differences along the $z$-axis are known. Moreover, the intensity is often uniformly split between different planes. Therefore, the localization task is finding emitter parameters (coordinates and intensity) in only one plane using data from all the planes, which in turn breaks the PSF symmetry and provides high axial precisions. 

The PSF engineering approaches attain precise axial localization of emitters by inscribing axial information in the shape of the PSF by simple wavefront modifications. The modification in wavefront can be achieved by inserting extra optical components \cite{huang2008three,baddeley2011three,izeddin2012psf} or computer controlled phase modulators \cite{pavani2009three,shechtman2014optimal,jia2014isotropic,shechtman2015precise} in the optical path. Several engineered PSFs have been devised for 3D super-resolution microscopy including an astigmatic PSF \cite{huang2008three}, double-helix PSF \cite{pavani2009three}, phase ramp PSF \cite{baddeley2011three}, corkscrew PSF \cite{lew2011corkscrew}, self bending PSF \cite{jia2014isotropic}, and tetrapod PSF \cite{shechtman2014optimal,shechtman2015precise}, Fig. \ref{Fig_PSF}. These PSFs have special characteristics, such as relative motions of different parts, width variations in different directions that quickly changes as a function of axial location, Fig. \ref{Fig_PSF}. The engineered PSFs cover different ranges of axial locations. The tetrapod PSF was designed to accomplish precise lateral emitter locations over the largest axial range compared to the other PSFs \cite{shechtman2014optimal,von2017three,sage2019super}. 

Most of the 2D emitter fitting algorithms discussed in the previous sections can be adapted for 3D emitter fitting \cite{sage2019super,zhou2019advances}. These approaches employ either a theoretical 3D PSF \cite{huang2008three,smith2010fast,thomann2002automatic,huang2017super,zhang20153d,huang20153d,mondal2016clean} or an experimentally acquired numeric PSF to calculate the model \cite{henriques2010quickpalm,juette2008three,grover2012super,liu2013three,babcock2017analyzing,li2018real,quirin2012optimal,min20143d,li2019depth,bohrer2018improved,ovesny2014high,shuang2016generalized,yi20193d}. More complicated PSFs such as tetrapod or double-helix have very intricate features that cannot be exactly expressed by an analytical function, so the PSF is generated using empirical data from calibration experiments. In the following, we discuss the PSF calibration procedure.

Data for PSF calibration is often a stack of images acquired at different $z$-locations by placing a fluorescent emitter at the origin of the lateral plane, $(x_0=0,y_0=0)$, and incrementing its axial location below and above the focal plane.
At far distances from the emitter, the PSF at the $z_i$-plane in the stack is given by the scalar diffraction theory \cite{goodman2005introduction,liu2013three}
\begin{equation}
\mathrm{PSF}(x,y;z_i) = \Bigg|\iint \left(\mathcal{P}(k_x,k_y)e^{2\pi ik_z(k_x,k_y)z_i}\right) e^{2\pi i\left(k_xx+k_yy\right)}dk_x\,dk_y\Bigg|^2
\label{Eq1_29} 
\end{equation}
where the frequency cutoff, $k_{\mathrm{max}}$, due to the microscope objective (see Fig. \ref{Fig_Res}), has to be considered in performing the integration. The above integral gives the inverse Fourier transform of the term inside the parentheses, which quantifies two effects that contribute to the wavefront distortion and thus deformation in the PSF shape including: 1) the pupil function, $\mathcal{P}=\mathcal{A}e^{i\Phi}$, is due to the scattering and attenuation of light traveling via different media such as through the sample itself ($\Phi$ can also be due to the phase intentionally added for PSF engineering); 2) the phase due to the axial ($z$) location of the emitter given by $k_z z_i$, where $k_z^2 = {n^2}/{\lambda^2} - (k_x^2 + k_y^2)$. $n$ and $\lambda$ are, respectively, the refractive index of the medium and the wavelength of light in vacuum. 

The camera only records the square magnitude of the above integral (which is the light intensity), while the phase information is mostly lost. However, given the stack of intensity PSF images acquired at multiple $z$-locations, the phase $\Phi$ can still be found employing phase retrieval techniques such as the Gerchberg-Saxton algorithm \cite{liu2013three,hanser2004phase,gerchberg1972practical} and the MLE approach \cite{aristov2018zola}. Substituting the obtained phase in eq. (\ref{Eq1_29}), one is able to obtain the PSF at any given $z$-location. However, the integral (\ref{Eq1_29}) is computationally expensive and therefore the PSF model is generated at just a few $z$-slices. The PSF at a desired location can then be numerically produced via either linear or spline interpolation of appropriate generated $z$-slices \cite{liu2013three,li2018real}. 
\subsection{Drift Correction}
\label{Sec1_5_5}
Drift is a common problem in super-resolution microscopy procedures, in which the sample alters its location overtime, resulting in distortion and degradation in the quality of final images. SMLM microscopy reconstructs high-quality images from a list of localizations collected over the course of an experiment, where even slight disturbances in the experiment lead to serious defects in the results. For instance, mechanical vibrations of the microscope stage or fluctuations in temperature result in rotational or translational movements of the sample during data acquisition. Since such disturbances are unavoidable, algorithms are required to measure and correct for drift in the image processing step. 

Multiple drift correction algorithms have been employed in super-resolution microscopy, such as use of fixed fiducial markers during the experiment \cite{betzig2006imaging,pertsinidis2010subnanometre,lee2012using,ma2017simple,colomb2017estimation,youn2018thermal,balinovic2019spectrally}, image cross-correlation between frames at different times \cite{mantooth2002cross,cizmar2011real,mlodzianoski2011sample,parslow2014sample,sugar2014free,tang2014sub,wang2014localization,smirnov2017automated}, computing drift directly from the list of found emitter coordinates \cite{geisler2012drift,elmokadem2015optimal,yothers2017real,wester2021robust,cnossen2021drift} and other procedures \cite{petersen2013image,marturi2013fast,qiu2013drift}. In what follows, we briefly introduce a few of these drift correction methods.

Fiducial markers are fixed point sources of light or structures during data acquisition that are used as reference points for drift correction. Since fiducial markers are fixed, their movements can be measured in the localization step and used to eliminate drift errors. Fiducial markers can introduce light corruption to the sample and some fiducial markers emit photons for a limited time, which restricts the data acquisition time. Cross-correlation approaches are the most common procedure for removal of drift error and different variants have been reported in the literature. Image noise deteriorates performance of the cross-correlation approaches to calculate drift errors. One algorithm maximizes the cross-correlation of the first frame with the rest of the sequence to calculate drift \cite{mlodzianoski2011sample,tang2014sub}. A fast implementation of cross-correlation is achieved in the Fourier domain \cite{parslow2014sample,sugar2014free,smirnov2017automated}, and some other approaches employ the cross-correlation of the sum images to reduce the effect of noise \cite{cizmar2011real,wang2014localization}. Some approaches have been developed using the list of found localizations to estimate sample drifts, such as the NND distribution of locations \cite{yothers2017real,wester2021robust,cnossen2021drift}.  

Wester, et al. \cite{wester2021robust} made use of a combination of the image registration approach and NND distribution of localizations to measure drift in the sample. The algorithm employs periodic 3D registration of the sample using brightfield images to remove drift errors. However, brightfield registration is only accurate to around $10 \, \mathrm{nm}$ and there might be still residual drift remaining. This approach thus uses post-processing of the localizations to extract residual drift from the NND distribution of emitter coordinates. This procedure is robust and also capable of calculating axial drift.   

In a recent work \cite{cnossen2021drift}, drift correction is performed using an entropy minimization scheme.  The SMLM reconstruction is considered to be a probablility distribution about the true emitter locations, and an entropy metric expressing the uncertainty in the position estimation is used as an optimization cost.  To simplify the calculation, the cost
is replaced by a closed form upper bound and points beyond a fixed distance threshold in an estimated covariance matrix are excluded.  The drift model is taken to be a piecewise cubic spline over a series of dataset segments. This algorithm outperforms the cross-correlation procedures.
\subsection{Fitting Quality and Image Quality}
\label{Sec1_5_6}
The quality of a fit is defined as how close the recovered location of an emitter is to its true location. 
The fitting quality depends on the emitter detection, emitter fitting and thresholding stages. Each of these steps were discussed in previous sections. There are a few metrics to assess the fitting quality including the Jaccard index (JAC), root mean square errors (RMSE) or accuracy, precision \cite{sage2015quantitative,sage2019super}, a vectorial model for localization uncertainties \cite{mazidi2018minimizing}, and a fit confidence test that examines the stability of a localization by exploring the model in the vicinity of the maximum likelihood \cite{mazidi2020quantifying}. 

Image resolution is often used as a reporter of image quality. However, there is no widely accepted definition for resolution in super-resolution microscopy literature \cite{deschout2014precisely,baddeley2018biological,demmerle2015assessing}. There are however a few common metrics that are used to assess microscopy image resolution such as localization precision \cite{deschout2014precisely}, SNR, Fourier ring correlation (FRC) \cite{nieuwenhuizen2013measuring}, and parameter-free resolution estimate \cite{descloux2019parameter}. In what follows, we discuss some of these metrics.   
\subsubsection*{Fitting Quality}
\begin{figure}[H]
\centering
\includegraphics[scale=1.7]{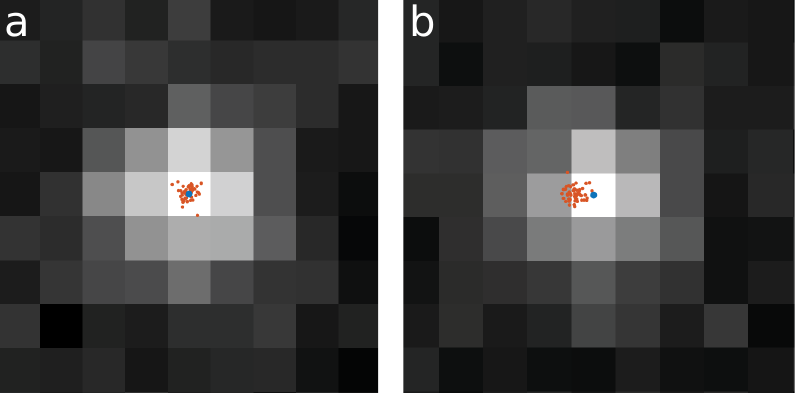}
\caption{Precision and accuracy concepts. Blue dots are the true emitter position and the orange dots stand for found locations. (a) Distribution of the found locations from an optimal fitting algorithm. Precision is proportional to the size of the distribution of the found locations. (b) Distribution of the found locations from a biased fitting algorithm. Accuracy is described as the deviation of the center of the distribution of found locations from the true position.}
\label{Fig1_11}
\end{figure}

The uncertainty in the location of an isolated emitter is characterized by precision and accuracy. Assuming that an emitter with location $x_0$ blinks multiple times, there are then multiple corresponding found locations, $x_f$, for the emitter. For an optimal fitting algorithm and under ideal conditions, the found locations are still slightly different and spread over a small region around the emitter location, $x_0$. The localization precision, $\sigma_{x}$, is the standard deviation of the distribution of $x_{f\,}$. This is a fundamental limit for localization precision imposed by the random nature of photons, and is not a result of instrument imperfections or flaws in the experiment design \cite{thompson2002precise,ober2004localization,deschout2014precisely}. Due to the stochastic nature of photons, each time that the emitter turns on, the number of photons reaching a certain pixel deviates by a small amount, which is called shot noise, resulting in small variations in the found emitter locations, Fig. \ref{Fig1_11}. The fundamental limit of precision for an optimal fitting algorithm is given by the square root of the inverse of the diagonal elements of the Fisher information matrix \cite{deschout2014precisely}
\begin{equation}
\sigma_{\theta_i} = \frac{1}{\sqrt{I\int_{-\infty}^{+\infty} \frac{1}{L(D|\theta)} \left( \frac{\partial L(D|\theta)}{\partial \theta_i} \right)^2 d\theta }}
\label{Eq1_30}
\end{equation}

\noindent
where $L(D|\theta)$ is the likelihood given in (\ref{Eq1_15}), $\theta, \theta_i$ and $I$ stand for the collection of all the parameters, the $i$th parameter, and the number of photons, respectively. The fundamental limit of localization precision for lateral coordinates due to shot noise is given by eq. (\ref{Eq1_6}). 

Accuracy describes the deviation of the average of the found locations from the true emitter location, Fig. \ref{Fig1_11}. Unlike localization precision, localization accuracy does not have a fundamental limit and can be zero \cite{deschout2014precisely}. However, due to instrument imperfections, optical aberrations and noise, the resultant localizations are often a little biased \cite{deng2009effect,mcgorty2014correction,coles2016characterisation,liu2018observing}. A set of acceptable localizations of an emitter have similar spreads of precisions and accuracies.

In practice, it is often challenging to achieve unbiased localizations and localization precisions close to the square root of the CRLB variance derived from a too simplistic model that only takes into account Poisson noise. This is due to other existing factors in super-resolution experiments such as  PSF mismatch, equipment (for instance cameras), and background noise. In what follows, we explain each of these effects:\\ \\
\textbf{PSF mismatch.}
Error-free PSF input to the iterative approaches is of key importance for accurate and precise emitter localization \cite{stallinga2010accuracy}. There are several factors that lead to major deviations between the expected PSF and the resultant PSF, including dipole orientation, emitter motions, and changes in the optical properties of the environment like the refractive index \cite{small2014fluorophore,deschout2014precisely,abraham2009quantitative,mortensen2010optimized,liu2018observing,liu2020three}. Fluorophore emitters are electric dipoles in nature and do not emit photons in an isotropic manner, despite the assumption of spatial uniform distribution of photons. A rigid emitter therefore has a nonuniform photon distribution which degrades the fitting quality when using an isotropic approximation of the PSF \cite{engelhardt2010molecular,stallinga2010accuracy,enderlein2006polarization}. However, these emitters are usually free to randomly rotate, which provides an isotropic distribution of photons. Another cause of PSF mismatch is emitter movements that exist in some experiments due to the nature of the sample, like live cell imaging or particle tracking, or due to bad fixation of the sample. This can result in a smearing pattern in the PSF known as the motion blur effect, specifically for large exposure times and fast diffusions, when the emitter emits photons while moving \cite{wong2010limit,deschout2012influence,michalet2012optimal}. Furthermore, imaging cellular structures within their native environment often requires light traveling through thick tissues of the specimen. The optical properties of the sample, such as the refractive index, changes as a function of location which leads to a disturbance in the light wavefront and therefore distortion of the PSF \cite{liu2018observing,liu2020three,mlodzianoski2018active,xu2020three}.      \\ \\
\textbf{Camera.} 
Several properties of cameras, such as noise introduced in the detection process and pixelation, affect the localization precision and accuracy. Two major detection noises introduced in cameras are read-out noise and the noise added in the amplification process that can affect localization uncertainty, see section 3. Model (\ref{Eq1_14}) does not take into account the detection noises, but can be modified to account for these noises by subtracting the read-out offset from the pixel values (input data) and rescaling them by the amplification factor in EMCCD cameras, where the read-out offset and amplification are uniform across the field of view. However, the detection noises require a more careful consideration in sCMOS cameras where these noises are unique for each pixel \cite{saurabh2012evaluation,huang2013video}. The pixel dependent noise in sCMOC cameras gives rise to a complicated and computationally expensive noise model. As such, in an effort to reduce the computational complexity, Huang et al. showed that this model can be approximately by a Poisson distribution with an additional source of photons \cite{huang2013video,liu2017scmos}.

Scientific cameras are not able to record the exact location of photons reaching them. However, they report the number of photons reached on a region within a certain area called a pixel. This causes loss of accurate locations of photons from an isolated emitter, and therefore results in localization uncertainty. This effect is called pixelation and is often negligible for relatively high SNR \cite{thompson2002precise,chao2013ultrahigh}. Moreover, sensitivity of cameras can be nonuniform over the field of view, which is another potential source of deterioration of localization precision \cite{deschout2014precisely}. \\ \\
\textbf{Background.}
Another limiting factor for precision and accuracy in localization microscopy is the background noise. Failure in correct background estimation can lead to biased localization estimates as was discussed in section 4.1. The homogeneous background is usually modeled by an additive offset to the intensity described in eq. (\ref{Eq1_13}). However, the situation is worse in the presence of structured background, which is more difficult to model and can give rise to more uncertainty in localizations. \\ 

We have so far discussed fitting quality for isolated emitters. However, overlapping PSFs from multiple active emitters in a diffraction limited area are unavoidable in SMLM. JAC is therefore introduced as a standard metric to measure the quality of fitting for a group of emitters in a dense region with overlapping PSFs \cite{sage2015quantitative,sage2019super,fazel2019bayesian}. JAC evaluates the rate of emitter detection versus the density of emitters. This is particularly important for multiple-emitter fitting algorithms which are supposed to detect and localize emitters in dense regions of data. To perform this test, sets of SMLM data with different emitter densities are simulated and processed with the given fitting algorithm. Next, the pairs of matching emitters within the sets of found and true locations are discovered by minimizing the cost metric between the two sets. JAC is then calculated as the ratio of the number of matched emitters (ME) to the total number of localizations within both sets
\begin{equation}
\mathrm{JAC}=\frac{\mathrm{ME}}{\mathrm{FE+TE}}
\label{Eq1_31}
\end{equation}      
where $\mathrm{FE}$ and $\mathrm{TE}$ refer to the number of found emitters and true emitters, respectively. Furthermore, Ram, et al. provided an approach to assess the quality of fitting for two nearby emitters employing Fisher information theory \cite{ram2013stochastic}.

Finally, theoretical inspection of fitting quality considering all the important factors is a difficult task, however statistical and experimental techniques have been proposed to calculate localization uncertainties taking into account all those factors \cite{endesfelder2014simple,lee2012double,tahmasbi2015determination}. An approach has been proposed that prescribes a simple way to compute localization precision from NND distribution of found localizations \cite{endesfelder2014simple}. Furthermore, Lee, et al. used well isolated emitters in sparsely labeled biological samples to experimentally measure localization precision under realistic experimental conditions \cite{lee2012double}. 
\subsubsection*{Image Quality}
After conducting an experiment and performing the image analysis, the final result is a super-resolved image reconstructed from the list of found localizations. What a scientist eventually cares about is the quality/resolution of this image and the amount of details that it reveals. The quality of the list of fits used to reconstruct an image directly affects the image resolution. For instance, the localization precisions must be smaller than a desired resolution. However, although the fitting quality is an essential factor in assessing the resolution of SMLM reconstructed images, it does not provide a comprehensive assessment of resolution for images of complex biological structures. For instance, an important factor in evaluating the resolution of reconstructed images of such complex structures is labeling density \cite{burgert2015artifacts,gould2012optical}. Furthermore, resolution highly depends on the type of structure being examined and differs from experiment to experiment \cite{deschout2014precisely,baddeley2018biological,venkataramani2016suresim}. For example, visualization of fine actin filaments requires uniform and high labeling densities \cite{riedl2008lifeact,desmarais2019optimizing}, while in imaging nuclear pore complexes (NPCs), the localization precision is of more importance since the sample itself has a discrete structure \cite{thevathasan2019nuclear}. To obtain a desired resolution, the labeling density should result in a NND of at least half of the desired resolution between the localized emitters according to the Nyquist-Shannon sampling theorem \cite{deschout2014precisely,nyquist1928certain,shannon1949communication,shroff2008live,fitzgerald2012estimation}.
Therefore, more advanced theoretical and experimental methods for evaluation of resolution in localization microscopy have been devised.
Here, we first provide a concise discussion on the labeling issue and then introduce some of the theoretical and experimental metrics for assessing the resolution.  

Low labeling density can be due to several problems in super-resolution microscopy, such as insufficient labeling, or broken labels or filtering localizations in the thresholding stage because of inadequate fitting quality. For some cases, such as experiments where there are multiple identical structures, this problem can be overcome by aligning and fusing the images of similar structures \cite{stathaki2011image,bates2018single,henderson2013avoiding}. Several alignment and fusion algorithms have been reported in the literature, including a few that require a template or structural assumptions in the alignment step \cite{loschberger2012super,mennella2012subdiffraction,szymborska2013nuclear,broeken2015resolution,shi2019deformed} as well as template-free particle fusion algorithms \cite{heydarian2018template,sieben2018multicolor,heydarian2019three,heydarian20213d}. 

A range of theoretical methods for evaluation of resolution in SMLM reconstructed images have been developed
such as the information transfer function (ITF) \cite{mukamel2012unified}, Fourier ring correlation (FRC) \cite{nieuwenhuizen2013measuring,banterle2013fourier}, super-resolution quantitative image rating and reporting of error locations (SQUIRREL) \cite{culley2018quantitative}, parameter-free resolution estimation \cite{descloux2019parameter}, and others \cite{cohen2019resolution,Marsh2021Sub}. In the following, we describe a few of these methods.\\ \\
\textbf{ITF.} 
Mukamel, et al. introduced the notion of ITF that accounts for photon statistics in localization microscopy as well as the resolution limit due to objective lenses. ITF is a generalization of the concept of MTF (see the Introduction) and reduces to the square of the MTF for conventional microscopy \cite{mukamel2012unified}. The ITF formalism gives image resolution bounds by providing a minimum on precision for image frequency estimates
\begin{equation}
\Delta_{I(k)} \leq F^{-1}(k)
\label{Eq1_35} 
\end{equation}
where $\Delta_I, k$ and $F$  stand for uncertainties, spatial frequency, and the ITF, respectively. This idea of resolution equally applies to both conventional and stochastic microscopy and unifies the concept of resolution \cite{mukamel2012unified}. Although ITF provides a generalized concept of resolution, it requires models of the target structure which makes it limited in practice \cite{deschout2014precisely}.\\ \\
\textbf{SQUIRREL.}
SQUIRREL builds an intensity image by convolving an input super-resolution image with a resolution scaling function. It then compares the obtained intensity image with the wide-field counterpart of the input super-resolution image to identify artifacts and disappearance of details \cite{culley2018quantitative}. SQUIRREL does not require any prior knowledge of the sample and is able to identify common super-resolution artifacts. However, out-of-focus light affects the performance of SQUIRREL and moreover, it cannot recognize small-scale artifacts. This approach returns a quantitative map of image anomalies and artifacts rather than reporting a resolution measure. The returned map can be used to optimize the experiment and image analysis, and improve the reconstructed image qualities. SQUIRREL was used to show that high image resolution does not necessarily correlate with artifact-free reconstructions \cite{culley2018quantitative}. \\ \\
\textbf{FRC.}
The FRC algorithm was adopted from electron microscopy \cite{van2000single} and adapted for resolution measurement in stochastic fluorescent microscopy \cite{nieuwenhuizen2013measuring,banterle2013fourier}. This approach uses two images from two identical structures or reconstruction of two images of the same structure by splitting the list of localizations into two halves. It then calculates the cross-correlations of the two images at different image frequencies, where components with frequency $k$ are given by a ring with radius $|k|$ in the Fourier domain. The higher frequencies represent finer details in the image. It is there the two images start to deviate and hence their cross-correlation rapidly declines. The frequency at which the cross-correlation falls below a given threshold can be utilized to calculate the image resolution. The method takes into account both localization precision and labeling density. It however has to be used with care for structures with more complicated geometries and discontinuous boundaries \cite{deschout2014precisely,baddeley2018biological}. \\ \\
\textbf{Parameter-free resolution estimate.}
Parameter-free resolution estimation \cite{descloux2019parameter} is based on the idea of correlation in the frequency domain similar to FRC. It however only uses a single image and furthermore does not require any threshold. The core idea is that structure information is mostly preserved in lower frequencies while high frequencies mostly represent noise. Therefore applying a frequency mask with a decreasing size will remove the contribution of noise and the correlation will gradually increase to a maximum in a certain frequency cutoff. After this frequency, the mask eliminates the structure information and the correlations falls. The frequency at which the maximum occurs is then employed to calculate the resolution \cite{descloux2019parameter}.\\

Now we turn to experimental approaches for characterization of resolution reported in the literature. For instance, DNA-rulers, DNA-origami structures, or standard biological structures such as NPCs with known spacing between emitters, have been employed as references to benchmark the ability of microscopy techniques in resolving closely spaced emitters \cite{thevathasan2019nuclear,steinhauer2009dna,raab2018using}. Some other well characterized biological structures such as microtubules and synthesized SMLM data for different structures, such as the Siemens star and crossing lines, have also been employed to inspect the resolution of SMLM reconstructed images \cite{sage2015quantitative,sage2019super,fazel2019bayesian,huang2011simultaneous,marsh2018artifact}.

Finally, resolution depends on multiple experimental and image analysis factors. A proper choice of labels, buffers and fixation protocols can lead to better localization uncertainties, less artifacts and therefore better image resolution. The choice of a localization algorithm also helps in optimizing both localization precision and localization density, reducing the image artifacts. The image rendering and visualization techniques also influence the image resolution. It has been demonstrated that different visualization procedures yield different resolutions for images reconstructed from the same list of emitter coordinates \cite{baddeley2010visualization}. 
\section{Applications of Single Molecule Localization Microscopy}
\label{Sec1_6}
Super-resolution microscopy has had significant contributions in shedding light on numerous biological problems since its advent in the 1990s. The major improvement in resolution of microscopy images helped with  unraveling details of many biological structures and processes, such as nuclear pore complexes \cite{thevathasan2019nuclear,heydarian2019three,schlichthaerle2019direct}, actin structures \cite{riedl2008lifeact}, cell membrane \cite{lingwood2010lipid,cambi2014cell}, cell division \cite{biteen2012three}, neurons \cite{dani2010superresolution,igarashi2018new}, plant cell biology \cite{fitzgibbon2010super,schubert2017super}, live cell imaging \cite{shroff2008live,westphal2008video,kner2009super}, and many more \cite{fornasiero2015super,sahl2017fluorescence,schermelleh2019super,vangindertael2018introduction,sauer2017single,kumar2018mechanism}. 
  
In addition to visualization of intricate structures, a tremendous quantitative insight into cell biology can be gained through post-processing of resulting localizations from SMLM techniques. These post-processing algorithms include single particle tracking (SPT) \cite{von2017three,shen2017single,manzo2015review}, cluster analysis \cite{griffie2016bayesian,nicovich2017turning}, molecule counting \cite{veatch2012correlation,lee2012counting,puchner2013counting,durisic2014single,rollins2015stochastic,fricke2015one,hummer2016model,karathanasis2017molecule,nino2017molecular} and particle fusion \cite{heydarian2018template,stathaki2011image,bates2018single,heydarian2019three,shi2019deformed,henderson2013avoiding,loschberger2012super,szymborska2013nuclear,broeken2015resolution,mennella2012subdiffraction,sieben2018multicolor}. In the following, SPT and clustering approaches are described in more detail.
\subsection{Single Particle Tracking}
\label{Sec1_6_1}
While super-resolution microscopy only deals with spatial resolution, Single Particle Tracking (SPT) is concerned with spatial as well as temporal resolution of dynamic samples. The microscopic world of biology is a live and vibrant place where many phenomena and interactive processes take place over time. An effective examination of these types of processes and interactions requires a comprehensive knowledge of dynamics and the motion of particles in those environments. 

SPT was first demonstrated by optical visualization of gold nano-particles on cell membranes \cite{geerts1987nanovid,de1988dynamic}, which paved the path for use of SPT in the study of cell membranes \cite{sheetz1989nanometre,saxton1997single}. Since the first demonstration of SPT in cell biology, there have been a burst of research using SPT that have revolutionized our understanding of the cell membrane \cite{lidke2004quantum,kusumi2005paradigm,alcor2009single,van2010nanometer}. An example of such research is the use of SPT in shedding light on the confined motion of receptors on the cell membrane imposed by the actin cytoskeleton \cite{kusumi2005paradigm,andrews2008actin}. SPT has also been employed to explore intracellular environments, including influenza and HIV virus internalization \cite{lakadamyali2003visualizing,brandenburg2007virus,ruthardt2011single}, and cargo transport along microtubules \cite{balint2013correlative}.          
  
SPT is comprised of three main steps: 1) detection and localization; 2) linking; and 3) post processing. The detection and localization of probes in SPT is similar to emitter localization in SMLM techniques and most of the fitting approaches discussed before can be employed for this stage of SPT. As described previously, more photons from the probes lead to more precise localizations. On the other hand, to increase the temporal resolution, a higher frame rate is desired during data acquisition. Higher frame rate means shorter exposure time and less photons and hence larger localization uncertainties. Therefore, bright probes are necessary for both high temporal resolution and precise localizations. Other important elements are photo-stability and size of the probes \cite{p2011probe}. Probes with long-lasting photo-stability and small sizes are desired for obtaining longer trajectories and less perturbation in the mobility and dynamics of the particles under study. There are three major categories of probes used in SPT: nano-particles, organic dyes and quantum dots (QDs) \cite{manzo2015review}. Nano-particles reflect light rather than emitting fluorescent light and have high photo-stability, but their size is rather large compared to typical proteins. Organic dyes are very small compared to typical proteins, but they are really dim and bleach rapidly \cite{manzo2015review}. The size of QDs are in-between organic dyes and nano-particles. They are relatively bright and photo-stable, which make them a perfect probe for most SPT experiments. QDs are manufactured with different emission wavelengths and are also suitable for multi-color SPT \cite{bruchez1998semiconductor,pinaud2010probing,cutler2013multi}.       

\begin{figure}[H]
\centering
\includegraphics[scale=0.65]{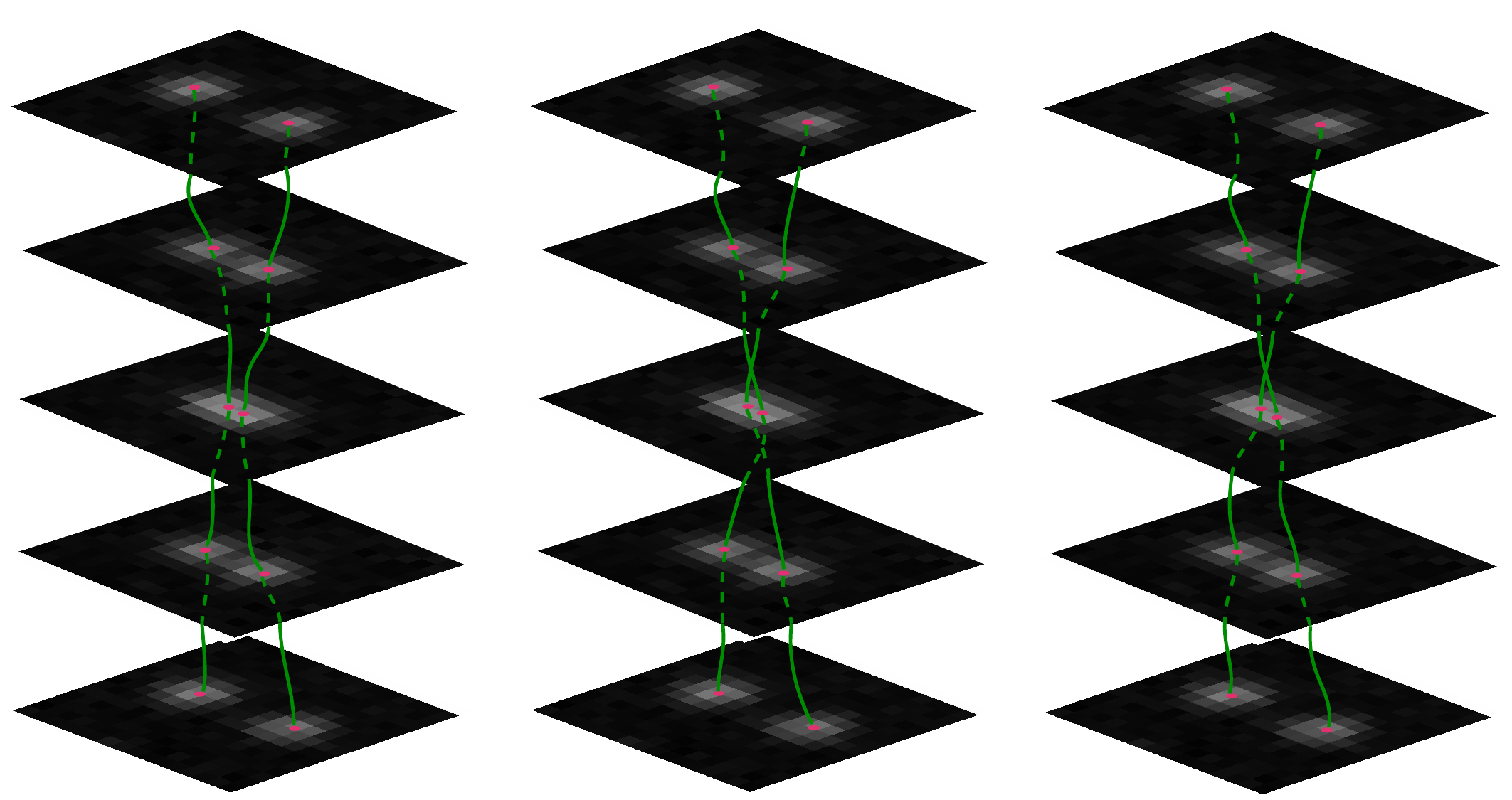}
\caption{Three possible links in the presence of two probes. The Red circles and green curves, respectively, represent detected probes and possible trajectories.}
\label{Fig1_13}
\end{figure}

After detection and localization, the detected probes are linked together across consecutive frames to construct particle trajectories. Linking is a non-trivial problem, even in the presence of only a few probes, Fig. \ref{Fig1_13}. Yet this is the case where there is an equal number of probes in every given frame, in the absence of blinking, bleaching, probes entering or leaving the field of view, and missed probes in the detection stage.  The reported linking algorithms in SPT can be categorized in two classes: 1) deterministic; 2) probabilistic; and 3) deep learning approaches \cite{newby2018convolutional,jakobs2019kymobutler,dmitrieva2019protein,granik2019single}. In what follows, we only describe deterministic and probabilistic techniques. Deterministic approaches give the same results every time, but probabilistic procedures use a random number generator at some point in the analysis and depending on the sequence of random numbers produced, they give slightly different answers each time. 

Given a set of localizations, deterministic approaches construct trajectories by linking the localizations so that the resulting trajectories minimize a certain cost 
function \cite{kalaidzidis2007intracellular}. The reported cost functions include distance \cite{ghosh1994automated,rogers2007precise,claytor2009accurately,applegate2011plustiptracker,patel2018rapid}, as well as concepts such as minimal energy path \cite{bonneau2005single,xue2009novel,lu2019minimal} or inertia \cite{sethi1987finding} borrowed from classical mechanics, and ad hoc cost functions that can be a function of parameters like probe location, intensity, and motion direction \cite{sage2003automatic,rink2005rab,sbalzarini2005feature,paavolainen2012application}. A range of different optimization algorithms have been employed to minimize the aforementioned cost functions, including the Hungarian \cite{applegate2011plustiptracker,rink2005rab,kuhn1955hungarian} and the greedy \cite{sethi1987finding,sage2003automatic,paavolainen2012application} methods. 

Probabilistic SPT approaches make use of a Bayesian algorithm to obtain the most probable tracks in the linking problem. SPT algorithms that implement a Bayesian procedure have demonstrated better performances compared to deterministic approaches due to inclusion of prior information \cite{chenouard2014objective}.

Here, we briefly describe the probabilistic approach. Assume a system with state $x_t$ at time $t$, which leads to an observation $z_t$, where $x_t \ne z_t$ due to noise. The state at time $t$ is a stochastic quantity related to the state at time $t-1$ as follows
\begin{equation}    
x_t \sim f_t(x_{t-1},u_t)
\label{Eq1_36}
\end{equation}
where $f_t$ and $u_t$ are, respectively, a function that characterizes the motion type and a variance. The observation also deviates from the state of the system due to noise and is given by 
\begin{equation}
z_t \sim g(x_t,v_t)
\label{Eq1_37}
\end{equation}
where $v_t$ is the noise variance. The objective is finding the state of the system at time $t$, $x_t$, given the list of observations up to this time, $z_{1:t}$. Using Bayes theorem, it can be shown that \cite{arulampalam2002tutorial}
\begin{equation}
P(x_t|z_{1:t})=\frac{P(z_t|x_t)P(x_t|z_{1:t-1})}{P(z_t|z_{1:t-1})}
\label{Eq1_38}
\end{equation}
where $P(z_t|x_t)$ is the likelihood given by the noise model (\ref{Eq1_37}), and $P(x_t|z_{1:t})$ is the posterior.
The prior is given by
\begin{equation}
P(x_t|z_{1:t-1})=\int P(x_t|x_{t-1})P(x_{t-1}|z_{1:t-1})dx_{t-1}
\label{Eq1_39}
\end{equation} 
where $P(x_t|x_{t-1})$ is defined by eq. (\ref{Eq1_36}).
Equations (\ref{Eq1_38}) and (\ref{Eq1_39}) provide a recursive formula to obtain the current state of the system from the list of observations up to this time along with the previous state of the system. However, these equations do not have an analytical solution for a general case. However, an analytical solution can be derived if the posterior has a Gaussian form at any time point, $f_t$ is a linear function and $g$ describes Gaussian noise, known as the Kalman filter \cite{ngoc1997adaptive,genovesio2003tracking,smal2008new,godinez2009deterministic,godinez2011tracking,godinez2014tracking,jaiswal2015tracking}. For instance, diffusion is a linear process and can be characterized by a Kalman filter. Another common filter is the particle filter \cite{smal2008multiple,arulampalam2002tutorial,khan2005mcmc,smal2006bayesian,godinez2007tracking,smal2008particle,yoon2008bayesian}, which is suitable for non-linear processes with non-Gaussian noise. 

The recursive method described above provides an approach that links new probes to the trajectories on a frame by frame basis. However, the Multiple Hypothesis Tracking (MHT) algorithm evaluates the possibility of different tracks, taking into account localized probes over either the entire data set or over a time window \cite{reid1979algorithm,cox1996efficient,jaqaman2008robust,chenouard2009multiple,chenouard2013multiple,liang2014novel}. While this procedure accomplishes the most optimal solution to the linking problem, it is computationally extremely expensive and impractical. Greedy algorithms accomplish approximations to the MHT optimal solution with tremendously less computational cost \cite{jaqaman2008robust,shafique2005noniterative}.

\begin{figure}[H]
\centering
\includegraphics[scale=1.65]{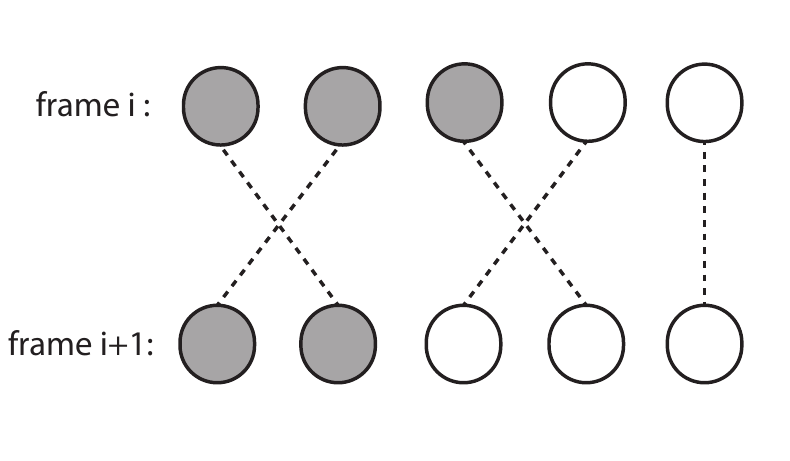}
\caption{An instance of probe linking in two consecutive frames in the presence of ghost probes. The blank and filled circles represent ghost and real probes.}
\label{Fig1_14}
\end{figure}

Jaqaman et al. \cite{jaqaman2008robust} proposed a relatively computationally inexpensive approximation to MHT using a greedy algorithm. This procedure uses a cost matrix approach to incorporate appearance (birth) and disappearance (death) of the probes in the model. They introduced ghost probes to consider three different linking scenarios when there is not an equal number of detected probes in consecutive frames, Fig. \ref{Fig1_14}. The real probes are connected together using a cost calculated from the localization precisions and kinetics of the probes
\begin{equation}
P(x_2,t_2|x_1,t_1)=\mathcal{N}(x_1,x_2,\sigma^2)
\label{Eq1_40}
\end{equation}
where $x$ stands for the probe locations and $\sigma^2$ is the variance due to diffusion and localization precisions, $\sigma_x$, defined as
\begin{equation}
\sigma^2=2D\Delta t + \sum_{i=1}^2\sigma_{xi}^2
\label{Eq1_41}
\end{equation} 
where $D$ and $\Delta t$ are the diffusion constant and exposure time, in turn.
A real probe can be connected to a ghost probe when a probe goes either off or on in a subsequent frame and the penalty is computed from the blinking statistics of the probes. The blinking process is assumed to be memoryless with $k_{\mathrm{on}}$ and $k_{\mathrm{off}}$ being the rate of the probes, respectively, going from off to on, and on to off
\begin{equation}
\begin{split}
P(\mathrm{\mathrm{off}}\rightarrow \mathrm{on}|k_{\mathrm{on}},\Delta t) =k_{\mathrm{on}} \Delta t \\
P(\mathrm{on}\rightarrow \mathrm{off}|k_{\mathrm{off}},\Delta t) =k_{\mathrm{off}} \Delta t.
\end{split}
\label{Equation1_42}
\end{equation}
Additionally, the probabilities for a probe staying on or off for $N$ frames is given as 
\begin{equation}
\begin{split}
P(\mathrm{Stay\,off}|N\Delta t)=P(\mathrm{Not\,going\,on}|N\Delta t)=(1-k_{\mathrm{on}}\Delta t)^N \\
P(\mathrm{Stay\,on}|N\Delta t)=P(\mathrm{Not\,going\,off}|N\Delta t)=(1-k_{\mathrm{off}}\Delta t)^N. 
\label{Eq1_43}
\end{split}
\end{equation}   
Eventually, ghost probes can also be linked together where the costs are given by the transpose cost matrix of the real probes. The Hungarian procedure is utilized to find the connections with the least cost. 

The first stage only connects probes across consecutive frames and produces short trajectories, Fig. \ref{Fig1_14}. In a second stage, the resulting short trajectories are connected together to obtain probe trajectories across the entire given sequence of the frames, which is called gap-closing. The gap-closing cost function is also a function of blinking statistics and localization precisions, except this time the probes are allowed to be linked together in non-consecutive frames. 

Once the trajectories are constructed, they can be employed to extract information about the underlying dynamics of the biological sample under study \cite{manzo2015review,qian1991single,michalet2010mean,michalet2012optimal,shen2017single}. Mean Square Displacement (MSD) is the most common post-analysis quantity of trajectories employed for this purpose. It is defined as the average of spatial distances between all pairs of localizations that are temporally separated by certain times, called the time lag. Given a trajectory comprised of $N$ localizations, the MSD can be calculated as follows:
\begin{equation}
\mathrm{MSD}(t_{\mathrm{lag}}=m\Delta t)=\frac{1}{N-m}\sum_{i=1}^{N-m}\left[x(t_i+m\Delta t)-x(t_i)\right]^2
\label{Eq1_44}
\end{equation}
where $x(t_i)$ is the location of the probe at time $t_i$ and $x(t_i+m\Delta t)$ is the probe location at $t_{\mathrm{lag}}=m\Delta t$ later. The MSD is a function of the time lag, $t_{\mathrm{lag}}$, and this function has been extensively investigated and well characterized for different motion types. By comparing the resulting MSD function from the obtained trajectories and the theoretical MSD forms, one is able to determine the underlying dynamic of the system under study. For diffusion or Brownian motion, the MSD is proportional to the time lag 
\begin{equation}   
\mathrm{MSD}(t_{\mathrm{lag}})=2dDt_{\mathrm{lag}}
\label{Eq1_45}
\end{equation}
where $d$ is the spatial dimensionality of the motion.
\subsection{Clustering}
\label{Sec1_6_2}
Clustering is the task of classifying a given set of data points into subsets in such a way that data points in a subset are more similar to each other than those in other subsets, based on a given property \cite{altman2017points}. Clustering has been employed in many scientific spheres, including localization microscopy, recognizing patterns and detecting communities. 

In SMLM, there are different levels of clustering for a list of localizations resulting from an experiment \cite{griffie2016bayesian,nicovich2017turning}. In SMLM, there are multiple observed localizations from an emitter over the course of data acquisition that form a cluster, which we will call here an intra-cluster. A higher level of clustering is clustering between emitters/molecules that signifies interactions between the molecules in a cluster at the biological level. We call this type of cluster inter-cluster. There is also co-clustering that takes place among different types of molecules imaged via multi-color super-resolution microscopy \cite{subach2009photoactivatable,spendier2012single,rossy2014method,georgieva2016nanometer}. 

Multiple algorithms have been developed or adapted to identify all sorts of clusters within SMLM localizations, including algorithms that inspect the extent of clustering within a data set  \cite{clark1954distance,zhang2006characterizing,owen2010palm,kiskowski2009use,sengupta2011probing,sengupta2012quantitative,khater2020review}, density based clustering algorithms 
\cite{khater2020review,ester1996density,dudok2015cell,rahbek2017super,andronov2016clustervisu,levet2015sr} and Bayesian clustering approaches \cite{richardson1997bayesian,griffie2016bayesian,khater2020review,rubin2015bayesian,post2019stochastic,fazel2019sub}. These approaches have been popular in post-processing of SMLM results of cell membrane samples, where it is conjectured that molecular clusters indicate underlying membrane structure or cell signaling \cite{williamson2011pre,owen2012sub,garcia2014nanoclustering,lin2016nanoscopic,chamma2018dynamics,roberts2018cluster}.   

\begin{figure}[H]
\centering
\includegraphics[scale=0.95]{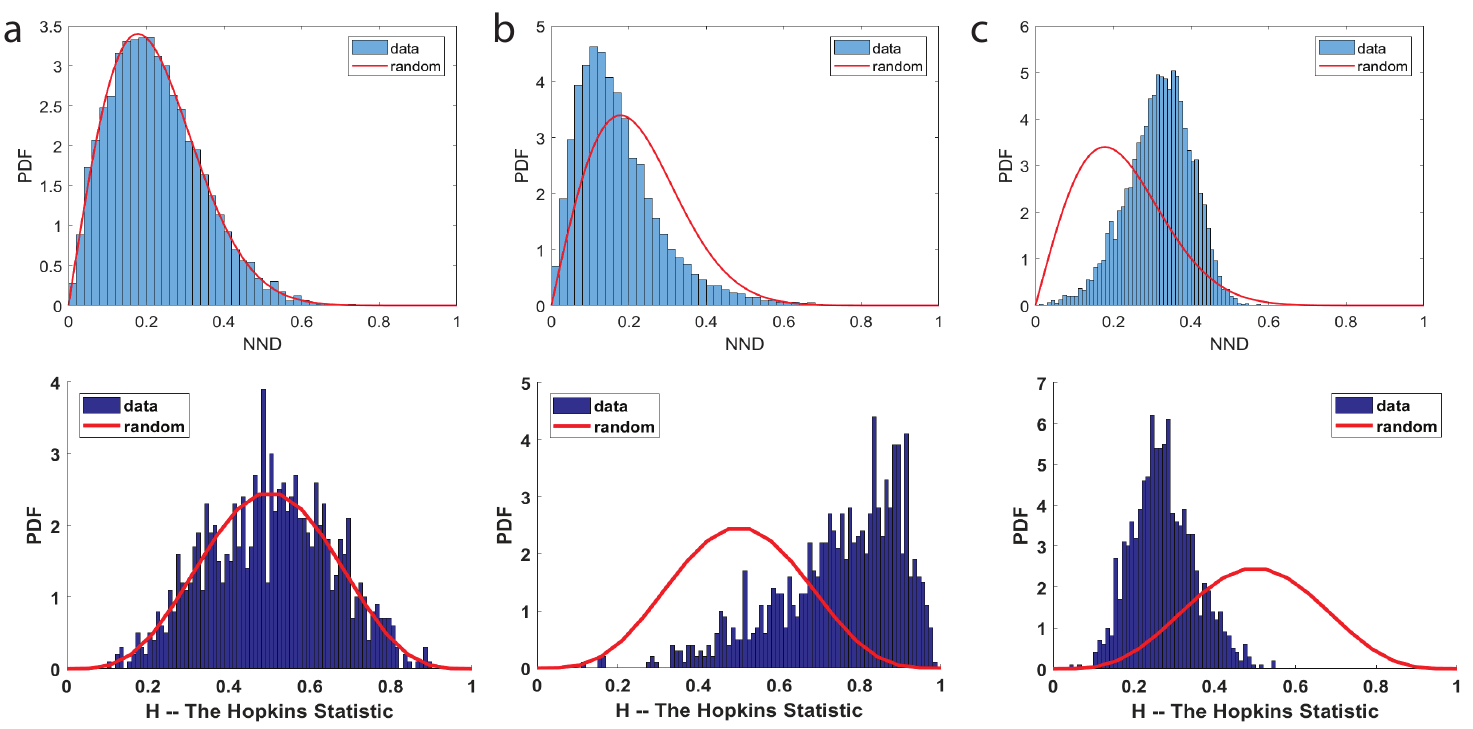}
\caption{NND distribution and Hopkins statistics. First and second rows represent the NND distribution and Hopkins statistics, respectively. (a) randomly distributed, (b) clustered, (c) more regularly spaced data.}
\label{Fig1_15}
\end{figure}

Some methods examine the degree of grouping or group properties of clusters, such as the shape or average density within a data set, Fig. \ref{Fig1_15}. A few of these approaches are NNDs \cite{clark1954distance}, Hopkins statistics \cite{zhang2006characterizing}, and Ripley's function and pair 
correlation \cite{owen2010palm,kiskowski2009use,sengupta2011probing,sengupta2012quantitative}. The NND probability distribution for uniform randomly distributed data is given by
\begin{equation}
f(r) = 2\pi \rho r e^{-\pi \rho r^2}
\label{Eq1_46}
\end{equation}
where $\rho$ is the density of the data points. The deviation of the NND distribution for a given data set from (\ref{Eq1_46}) indicates either clustering or a more regularly spaced data set, Fig. \ref{Fig1_15}. Hopkins statistics ($H$) tests for spatial randomness of a point pattern by comparing NNDs between a given data set and a uniform randomly distributed data set. Values of $H$ near $0.5$ imply randomly distributed data, while values near 1 indicate highly clustered data, and values near zero signify more regularly spaced data, Fig \ref{Fig1_15}. 

The Ripley's K-function, $K(r)$, is another statistical test that explores the extent of clustering within a given data set. Defining $k_i(r)$  as the number of points within distance $r$ from the $i$th point, $K(r)$ is given by the average of the $k_i$ for all data points divided by the mean density over all data points. The clustering behavior of the given data set is then given by
\begin{equation}
\begin{split}
K(r) < \pi r^2 \hspace{1cm} \mathrm{regularly \, spaced} \hspace{3mm}\\
K(r) = \pi r^2 \hspace{1cm} \mathrm{random} \hspace{1.52cm} \\
K(r) > \pi r^2 \hspace{1cm} \mathrm{clustered} \hspace{1.35cm}
\end{split}
\label{Eq1_47}
\end{equation}
Pair correlation analysis is the same idea as Ripley's K-function extended to images via pixels rather than a set of points.

A range of clustering algorithms have been developed that take advantage of density fluctuations within a data set to identify clusters. DBSCAN \cite{ester1996density} and Voronoi tessellation \cite{andronov2016clustervisu,levet2015sr} algorithms are among the most popular density based approaches in SMLM data post-analysis. DBSCAN has two inputs. One of them is the maximum distance between the points within a cluster and the other is the minimum number of data points within a cluster. The algorithm starts from a random point and adds points to the set until there is no more points within the given maximum distance. If the number of points inside the set is more than the given threshold, it will be returned as a cluster, otherwise it is considered as an outlier. Voronoi tessellation draws boundary line segments between every pair of data points in the given data set so that the distance from every member of a pair to the line between them is equal. The final geometry is comprised of polygons where each polygon contains only one data point, called a Voronoi diagram. This algorithm identifies clusters by inspecting the number of vertices and the area of the polygons at different neighboring levels.       

Bayesian clustering algorithms for SMLM data are superior to density based approaches because they take advantage of prior knowledge. Available prior information can be included in Bayesian algorithms via prior distributions. Rubin-Delanchy, et al. developed an algorithm to identify inter-clusters within SMLM localizations, called Bayesian cluster identification \cite{griffie2016bayesian,rubin2015bayesian}. This algorithm classifies the localizations into either background or signal by means of a binomial distribution. A large number of clustering proposals are then put forward using a procedure similar to Ripley's K-function. Scores for different clustering proposals are calculated using the posterior, where the Dirichlet distribution is the prior on different arrangements. At the end, the configuration with the highest score is returned as the solution. 

Post, et al. proposed a Bayesian clustering algorithm to identify intra-clusters (emitters) on the surface of spherical nano-particles \cite{post2019stochastic}. Their algorithm takes full advantage of localization precisions, but it does not return the number of emitters on the surface of every individual nano-particle. Otherwise, they find the mean number of emitters on the surface of nano-particles by fitting a log-normal distribution to the number of localizations per nano-particle. They next employ this piece of information to localize those emitters and then inspect their inter-clustering behavior using the NND distribution. 

Fazel et al. developed Bayesian Grouping of Localizations (BaGoL) that takes advantage of localization precisions as well as mean number of blinking/binding events per emitter to identify and localize emitters with sub-nanometer precision in dense regions \cite{fazel2019sub}. This algorithm uses NND distributions and an intensity filter to recognize and remove outliers.

Lin, et al. developed a non-Bayesian intra-clustering algorithm based on a hierarchical clustering paradigm, which allows exploitation of localization precisions \cite{lin2016nanoscopic}. This algorithm proposes different clustering configurations for a given set of localizations similar to the hierarchical procedure, and next performs a hypothesis test to pick one of the proposed clustering models.   
\section*{Conclusion and Future Outlook}
Two decades after the advent of super-resolution microscopy, these techniques discussed above have become standard tools in direct and non-invasive examination of sub-cellular environments with unprecedented resolution. Among different super-resolution techniques, the SMLM approach involves data processing and post-processing stages that are reviewed in this survey. Yet, despite all the progress in this area, the existing SMLM data analysis techniques face multiple challenges. These include the existence of sample induced optical aberrations, and the lack of common metrics in assessing the robustness of image analysis and image resolution. Furthermore, 
extracting meaningful biological information from the resulting list of SMLM emitter coordinates are complex and difficult tasks. Therefore, the area of SMLM data post-processing is a fast growing field opening many promising venues of research in this area which, in turn, demands new computational tools.

The 3D nature of biological structures has been a long time motivation for developing 3D SMLM microscopy techniques. Particularly, it has been of interest to observe subcellular structures within their native environments. However, fluorescent light traveling within such complex biological environments is significantly perturbed, degrading the quality of the final SMLM images. Adaptive optics \cite{liu2018observing} is an established experimental technique to compensate for such sample induced aberrations, but the potential of computational methods to deal with this major problem is still under-explored \cite{xu2020three}. Another main challenge in 3D SMLM is visualization of 3D structures, which demands the development of more sophisticated visualization tool-boxes able to robustly and rapidly handle millions of localizations. 

Another major challenge in SMLM data analysis is assessment of fitting and image qualities. In spite of numerous reported metrics and methods in the literature (covered in section 4.6), 
there is still no standardized universal set of metrics for comparing analysis results. For example, there can be disagreement between different existing metrics in evaluating the robustness of a fitting algorithm or the resolution of an image \cite{culley2018quantitative}. 

An obstacle in obtaining meaning from SMLM and SPT data is the development of
multi-color and multi-target analysis techniques that can identify spatial relationships between multiple species of proteins, and techniques for inferring protein-protein interactions \cite{low2011erbb1,wang2009quantitative,larson2014design,kumar2018mechanism}. The main issues in this area are image registration and colocalization between different channels. These are typically performed using different correlation techniques \cite{adler2010quantifying,dunn2011practical}, but such methods are often hindered by noise \cite{fletcher2010multi,lagache2015statistical}. The use of machine learning techniques seems to be a promising solution \cite{ordabayev2021bayesian}.

Solutions to the current SMLM data analysis challenges discussed here as well as many others might be found in leveraging tools from cutting-edge machine learning methods such as Bayesian techniques and neural networks. However, each of these paradigms comes with its own benefits and shortcomings. For example, Bayesian techniques provide robust analysis tools for data with low SNR, and takes into account different existing sources of uncertainty. Nonetheless, Bayesian techniques are often slow and require multiple input parameters. On the other hand, deep learning techniques are fast and parameter-free, but they do not provide any uncertainty estimate over the learned quantities and require careful training. A promising emerging data analysis avenue is Bayesian neural networks which addresses some of the mentioned issues by combining Bayesian techniques and neural networks. Finally, an important guideline in developing SMLM data analysis methods is the accessibility of these advanced machine learning tools to non-expert users by developing user friendly image processing packages and providing thorough algorithmic and implementation oriented (user) documentation. 

\section*{Acknowledgement}
We thank Keith A. Lidke and Steve Press\'e for useful comments.
\section*{Data Availability}
Data sharing is not applicable to this article as no new data were created or analyzed in this work.
\section*{Conflict of Interest}
Authors declare no conflict of interests.  
\newpage
\nocite{*}
\bibliography{Ref}
\bibliographystyle{unsrt}
\end{document}